\documentclass[aip,pof,preprint,12pt,a4paper,amsmath,amssymb]{revtex4-1}
\usepackage[english]{babel}
\usepackage{graphicx}

\renewcommand{\cite}[1]{[\onlinecite{#1}]}
\providecommand{\citetwo}[2]{[\onlinecite{#1},\onlinecite{#2}]}

\providecommand{\enlarge}[1]{}
\renewcommand{\enlarge}[1]{\typeout{WARNING: page enlarged}\enlargethispage{#1}}

\DeclareMathOperator{\const}{const}
\renewcommand{\vec}[1]{\ensuremath{\mathbf{#1}}}

\renewcommand{\subsection}[1]
	{\par\pagebreak[3]\refstepcounter{subsection}\textit{#1}\par}

\providecommand{\NPPS}{Nernst-Planck-Poisson-Stokes}

\begin{document}

\title{Direct Numerical Simulation of Electrokinetic Instability and~Transition
to Chaotic Motion}

\author{\firstname{E.~A.}~\surname{Demekhin}}
\email{edemekhi@gmail.com}
\affiliation{Laboratory of Micro- and Nanofluidics,\\
Moscow State University, Moscow, 119192, Russian Federation}
\affiliation{Department of Computation Mathematics and Computer Science,\\
Kuban State University, Krasnodar, 350040, Russian Federation}
\affiliation{Institute of Mechanics,\\
Moscow State University, Moscow, 117192, Russian Federation}

\author{\firstname{N.~V.}~\surname{Nikitin}}
\affiliation{Institute of Mechanics,\\
Moscow State University, Moscow, 117192, Russian Federation}

\author{\firstname{V.~S.}~\surname{Shelistov}}
\affiliation{Scientific Research Department,\\
Kuban State University, Krasnodar, 350040, Russian Federation}
\affiliation{Institute of Mechanics,\\
Moscow State University, Moscow, 117192, Russian Federation}

\begin{abstract}
\scriptsize
A~new type of instability --- electrokinetic instability --- and an unusual
transition to chaotic motion near a charge-selective surface (semi-selective
electric membrane, electrode or system of micro-/nanochannels) was studied by
numerical integration of the \NPPS\ system and a weakly nonlinear analysis near
the threshold of instability. A~special finite-difference method was used for
the space discretization along with a semi-implicit
$3\frac{1}{3}$\nobreakdash-step Runge-Kutta scheme for the integration in time.
The linear stability problem was resolved by a spectral method. Two kinds of
initial conditions were considered: (a) white noise initial conditions to mimic
``room disturbances'' and subsequent natural evolution of the solution; (b) an
artificial monochromatic ion distribution with a fixed wave number to simulate
regular wave patterns. The results were studied from the viewpoint of
hydrodynamic stability and bifurcation theory. The threshold of
electroconvective movement was found by the linear spectral stability theory,
the results of which were confirmed by numerical simulation of the entire
system. The following regimes, which replace each other as the potential drop
between the selective surfaces increases, were obtained: one-dimensional steady
solution; two-dimensional steady electroconvective vortices (stationary point in
a proper phase space); unsteady vortices aperiodically changing their parameters
(homoclinic contour); periodic motion (limit cycle); and chaotic motion. The
transition to chaotic motion did not include Hopf bifurcation. Numerical
resolution of the thin concentration polarization layer showed spike-like charge
profiles along the surface, which could be, depending on the regime, either
steady or aperiodically coalescent.

The numerical investigation confirmed the experimentally observed absence of
regular (near-sinusoidal) oscillations for the overlimiting regimes. There is a
qualitative agreement of the experimental and the theoretical values of the
threshold of instability, the dominant size of the observed coherent structures,
and the experimental and theoretical volt--current characteristics.
\end{abstract}



\maketitle

\section{Introduction}
\enlarge{1em}

\subsection{General background}
Problems of electrokinetics have recently attracted a great deal of attention
due to rapid developments in micro-, nano- and biotechnology. Among the numerous
modern applications of electrokinetics are micropumps, biological cells,
electropolishing of mono- and polycrystalline aluminum, and the growth of
aluminum oxide layers for creating micro- and nanoscale regular structures such
as quantum dots and wires.

The study of the space charge in the electric double-ion layer in an electrolyte
solution between semi-selective ion-exchange membranes under a potential drop is
a fundamental problem of modern physics, first addressed by Helmholtz. The first
theoretical investigation of the regime of limiting currents was done in the
works by Grafov and Chernenko \cite{GrCh1}, Smyrl and Newman \cite{Smr1} and
Dukhin and Deryagin \cite{DukhDe}. Rubinstein and Shtilman \cite{Rub1} described
the regime of limiting currents (see also \cite{Buk1,Nik1,Lis1,Man1,Bab5,Chu1}).
Hydrodynamics is not involved in either the underlimiting or limiting regime,
and both regimes are one-dimensional. A~curious electrokinetic instability and
pattern formation were theoretically predicted by Rubinstein and Zaltzman
\citetwo{Rub3}{Rub10} as a bifurcation to overlimiting regimes, which also makes
the solution two-dimensional. Rubinstein, Staude and Kedem \cite{Rub23} found
experimentally that the transition to the overlimiting regime was accompanied by
current oscillations. These facts indirectly indicated a correlation between the
instability and overlimiting currents. This correlation was also demonstrated in
Maletzki et al.~\cite{Mal}, Belova et al.~\cite{Belv}, and Rubinstein et
al.~\cite{Rub12}, where the overlimiting regimes were eliminated when
instability was artificially suppressed. The first direct experimental proof of
the electroconvective instability that arises with an increasing potential drop
between the bulk ion-selective membranes was reported by Rubinstein et
al.~\cite{RubRub1}, who managed to show the existence of small vortices near the
membrane surface. Yossifon and Chang~\cite{YCh} also observed an array of
electroconvective vortices arising under the effect of a slow alternating
electric field, while Kim et al.~\cite{KmHn} found a nonequilibrium vortices DC
field. A~unified theoretical description of electrokinetic instability, valid
for all three regimes, was presented by Zaltzman and Rubinstein \cite{Rub13},
based on a systematic asymptotic analysis of the problem. The numerical analysis
of the nonlinear regimes of the electrokinetic instability was developed in
\cite{DemShel,DemChang,Han1}. The self-similar character of the initial stage of
the evolution and instability of the self-similar solution were investigated in
\cite{DemShel,KD2,DemSh,Dem}.

\subsection{Electrokinetic instability}
\begin{figure}[hbtp]
\centering
 \includegraphics[width=\textwidth]{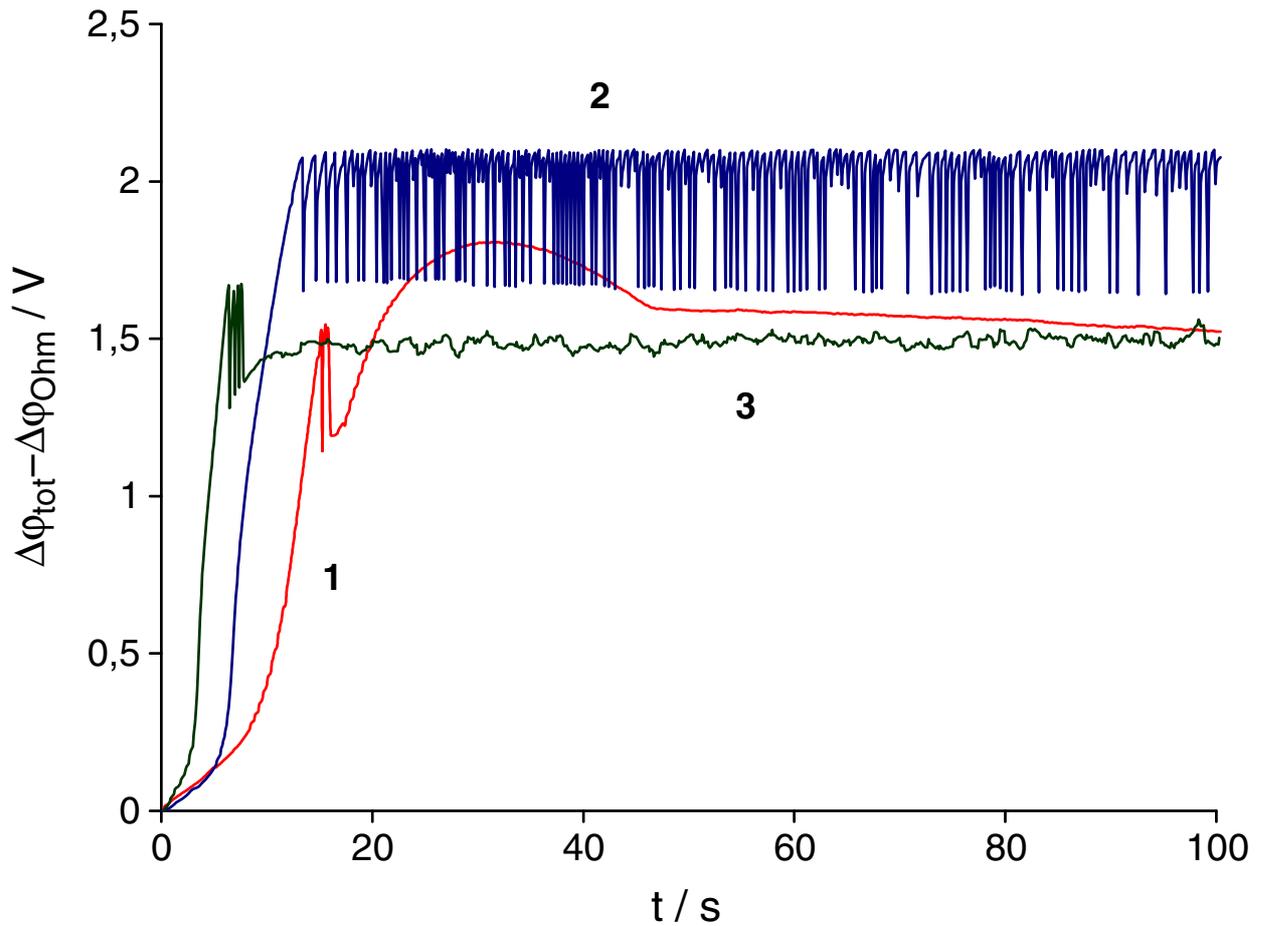}
\caption{(Color online) Chronopotentiograms time vs voltage in $NaCl$ solutions
at different ratios of the current density to its limiting value: 1~--- 1.5;
2~--- 2.5; 3~--- 3.6 \cite{VNic1}}
\label{VNic1}
\end{figure}

The fundamental interest in the problem is connected with a novel type of
electrohydrodynamic instability: electrokinetic instability. This instability
triggers a hydrodynamic flow and, in turn, intensifies the ion flux, which is
responsible for the overlimiting currents. Although the cells observed in the
electroconvective motion look like the cells in Rayleigh-B\'enard convection and
B\'enard-Marangoni thermoconvection, the electroconvective instability is much
more complicated from both the physical and the mathematical points of view. The
Reynolds number in the electrokinetic instability is very small and, hence, the
dissipation is very large and the nonlinear terms in Navier-Stokes system are
negligibly small. The nonlinearity responsible for bifurcation arises from other
equations of the system (\NPPS\ system). This explains the dramatic distinction
between the chaotic motion in the macro- and microhydrodynamics. The
electrohydrodynamics of the electrolytes between the membranes manifests new
types of bifurcation and transition to chaos, developed in both micro- and
nanoscales. In particular, the large dissipation of the system should suppress
its oscillatory motion, in particular, Hopf bifurcation. In Fig.~\ref{VNic1},
the results of experiments \cite{VNic1} do not reveal any of the regular
sinusoidal behavior that is characteristic near a Hopf bifurcation point.

\subsection{Numerical tools and methods of the present work}
During the last decades, direct numerical simulations have been recognized as a
powerful and reliable tool for studying many classical hydrodynamical problems,
in particular, for modeling laminar--turbulent transitions and chaotic turbulent
flows. It is then logical to apply this tool to study electrokinetic
instability. The system of the \NPPS\ equations has a small parameter, the Debye
number, at the highest derivatives. As a result, a thin space charge region with
a rapid changing of the unknowns is formed near the surface. This causes
significant difficulties for its numerical solution. These difficulties are
compounded by the complexity of the chaotic regime when the flow contains a wide
range of different scales. There are two approaches to overcoming these
difficulties. The first one is semi-analytical, when the solution in the
extended space charge region is sought analytically as the inner expansion but
numerically in the diffusion region which is treated as the outer expansion,
with a proper matching of the inner and outer expansions, see
\citetwo{Rub10}{Pun2}. The second approach solves the entire system of {\NPPS}
equations numerically, without any simplification. In this case, the extensive
experience of turbulent flow simulation for the Navier-Stokes equations can be
employed.

In our previous works, \citetwo{DemShel}{DemChang}, a quasi-spectral
discretization with Fourier series and Chebyshev polynomials was applied in
the spatial directions; this method happened to be computer time consuming. In
the present work, a special finite-difference scheme with a grid point
accumulation in the extended space charge region near the wall has been
developed.

Spatial discretization with fine resolution leads to stiff problems and requires
implicit methods for time advancement. Fully implicit methods produce a set of
nonlinear coupled equations for the problem variables on the new time level, and
are usually prohibitively costly for long-term calculations. Semi-implicit
methods, in which only a part of the operator is treated implicitly, constitute
a reasonable compromise for this class of problems.

A~semi-implicit three-step Runge-Kutta scheme with third-order accuracy in time
was adopted from \cite{Niki1,Niki2,NVN2}, where it was applied to the unsteady
Navier-Stokes equations. The scheme is supplied with a built-in local accuracy
estimation and time-step control algorithm. The idea of the time-step control
algorithm is not novel. It was first utilized in semi-implicit Runge-Kutta
schemes for the Navier-Stokes equations in \citetwo{Niki3}{Niki4}. Having been
extensively exploited for a period of about 10 years, it proved itself to be
efficient and convenient, especially for flows with wide variations in
characteristic time scales, for instance, in simulations of laminar--turbulent
transitions.

The results of the extensive numerical simulation are discussed and understood
from the viewpoint of bifurcation theory. Two kinds of initial conditions have
been considered: artificial monochromatic perturbations and white noise initial
conditions to mimic natural room disturbances. The context of the present work
is the set of attractors of a high dimensional dynamical system generated by the
original \NPPS\ equations. The averaged electric current and its time derivative
have been taken as a representative low dimensional projection to describe the
behavior of the whole system, its transitions, and attractors.

\subsection{Outline}
The present paper is structured as follows. In Part~2, the governing equations
for the dilute electrolyte, the \NPPS\ system, along with the relevant boundary
and initial conditions (of two kinds) are presented in dimensionless form and
typical characteristic values of the system parameters are evaluated.

Part~3 describes the details of the finite-difference scheme with respect to
both the spatial variables and time. The selection of the domain length and
discretization of the initial conditions are discussed at the end of Part~3.

Part~4 presents the results of the numerical simulation, and they are then
discussed from the viewpoint of bifurcation theory and hydrodynamic stability.
The potential drop $\Delta V$ is taken as a control parameter, and the
bifurcation points and corresponding regimes are investigated in terms of this
parameter. At the end of this part, the results of the simulation are
qualitatively compared with experiments.

Part~5 summarizes and discusses the research, and presents the main conclusions.

\section{The mathematical model}
A~symmetric, binary electrolyte with a diffusivity of cations and anions
$\tilde{D}$, dynamic viscosity $\tilde{\mu}$ and electric permittivity
$\tilde{\varepsilon}$, and bounded by an ideal, semi-selective ion-exchange
membrane surfaces, $\tilde{y}=0$ and $\tilde{y}=\tilde{L}$, is considered.
Notations with tilde are used for the dimensional variables, as opposed to
their dimensionless counterparts without tilde. $\{\tilde{x},\tilde{y}\}$ are
the coordinates, where $\tilde{x}$ is directed along the membrane surface and
$\tilde{y}$ is normal to the membrane surface.

The characteristic quantities to make the system dimensionless are:\\
\begin{tabular}{@{}l@{ --- }l}
$\tilde{L}$ & the characteristic length, the distance between the membranes;\\
$\tilde{L}^2/\tilde{D}$ & the characteristic time;\\
$\tilde{\mu}$ & the dynamic viscosity;\\
$\tilde{\Phi}_0=\tilde{R}\tilde{T}/\tilde{F}$ & the thermodynamic potential;\\
$\tilde{c}_0$ & the bulk ion concentration at the initial time.
\end{tabular}\\
Here $\tilde{F}$ is Faraday's constant, $\tilde{R}$ is the universal gas
constant and $\tilde{T}$ is the temperature in kelvins. In terms of the
dimensional variables, the flow is confined between rigid walls at $y=0$ and
$y=1$.

The electroconvection is described by the equations for the ion transport, the
Poisson equation for the electric potential, and the Stokes equations for a
creeping flow:
\begin{eqnarray}
\frac{\partial c^\pm}{\partial t} + \vec{u}\cdot\nabla c^\pm & = &
 \pm\nabla\cdot\left(c^\pm\nabla\Phi\right) + \nabla^2 c^\pm, \label{eq1}\\
\nu^2\nabla^2\Phi & = & c^--c^+ \quad \equiv -\rho, \label{eq2}\\
-\nabla\Pi + \nabla^2\vec{u} & = & \frac{\varkappa}{\nu^2} \rho \nabla\Phi,
 \quad \nabla\cdot\vec{u} = 0. \label{eq3}
\end{eqnarray}
Here $c^+$ and $c^-$ are the molar concentrations of cations and anions;
$\vec{u}=\{u,v\}$ is the fluid velocity vector; $\Phi$ is the electrical
potential; $\Pi$ is the pressure; $\nu$ is the dimensionless Debye length or
Debye number, which is the ratio of the Debye length $\tilde{\lambda}_D$ with
the macroscopic length $\tilde{L}$,
\[
\nu=\frac{\tilde{\lambda}_D}{\tilde{L}}, \qquad
\tilde{\lambda}_D=\left(\frac{\tilde{\varepsilon}\tilde{\Phi}_0}{\tilde{F}\tilde{c}_0}\right)^{1/2}=
\left(\frac{\tilde{\varepsilon}\tilde{R}\tilde{T}}{\tilde{F}^2\tilde{c}_0}\right)^{1/2},
\]
and $\varkappa$ is a coupling coefficient between the hydrodynamics and the
electrostatics,
\[ \varkappa=\frac{\tilde\varepsilon\tilde\Phi_0^2}{\tilde\mu\tilde{D}}. \]
It characterizes the physical properties of the electrolyte solution and is
fixed for a given liquid and electrolyte.

If the flow is two-dimensional (2D), instead of the system \eqref{eq3}, one
equation for the stream function $\Psi$, $(u,v)=(\partial\Psi/\partial y,
-\partial \Psi/\partial x)$, can be considered,
\begin{equation}\label{eq3a}
\nabla^4\Psi = \frac{\varkappa}{\nu^2}\left\{
 \frac{\partial}{\partial y}\left(\rho\frac{\partial\Phi}{\partial x}\right)
 - \frac{\partial}{\partial x}\left(\rho\frac{\partial\Phi}{\partial y}\right)
 \right\}.
\end{equation}

The above system of equations is complemented by the proper boundary conditions:
\begin{align}
y &= 0: & c^+ &= p, & \displaystyle -c^-\frac{\partial\Phi}{\partial y}
 + \frac{\partial c^-}{\partial y} &= 0, & \Phi &= 0, & \vec{u} &= 0; \\
y &= 1: & c^+ &= p, & \displaystyle -c^-\frac{\partial\Phi}{\partial y}
 + \frac{\partial c^-}{\partial y} &= 0, & \Phi &= \Delta V, & \vec{u} &= 0.
\label{eq5}
\end{align}

The first boundary condition, prescribing an interface concentration equal to
that of the fixed charges inside the membrane, is asymptotically valid for large
$p$ and was first introduced by Rubinstein (see, for example, \cite{Rub13} and
corresponding references) to avoid solution inside the membrane. The second
boundary condition means no flux for negative ions, the third condition is a fixed potential drop, and the last condition is that the velocity vanish at the
rigid surface. The spatial domain is assumed to be infinitely large in the
$x$\nobreakdash-direction, and the boundedness of the solution as
$x \to \pm \infty$ is imposed as a condition.

Adding initial conditions for the cations and anions completes the system 
\eqref{eq1}--\eqref{eq5}. These initial conditions arise from the following
viewpoint: when there is no potential difference between the membranes, the
distribution of ions is homogeneous and neutral. This corresponds to the
condition $c^+ = c^- = 1$. Some kind of perturbation should be superimposed on
this distribution,
\begin{equation}\label{eqq6}
t=0: \qquad c^+ = 1+\hat{c}^+(x), \qquad c^- = 1+\hat{c}^-(x).
\end{equation}

Two kinds of initial conditions \eqref{eqq6} are considered.

(a) The initial conditions, which are natural from the viewpoint of the
experiment. The so-called ``room perturbations'' determining the external
low-amplitude and broadband white noise should be imposed on the concentrations.
These initial conditions are
\begin{equation}\label{eqq67}
\hat{c}^+=\int_0^{\infty}\hat{p}(k) e^{-ikx}\,dk,
 \qquad \hat{c}^-=\int_0^{\infty}\hat{n}(k) e^{-ikx}\,dk.
\end{equation}

(b) Artificial forced monochromatic perturbations with a fixed wave number
$k_s$,
\begin{equation}\label{eqq66}
\hat{c}^+=\hat{p}\cdot\cos k_s x,
 \qquad \hat{c}^-=\hat{n}\cdot\cos k_s x.
\end{equation}

The characteristic electric current $j$ at the membrane surface is referred to
the limiting current, $j_{\lim}=4$,
\begin{equation}\label{eq777}
j = \frac{1}{4}\left(c^+\frac{\partial\Phi}{\partial y}
 + \frac{\partial c^+}{\partial y}\right) \quad \text{for} \quad y = 0.
\end{equation}
It is also convenient for our further analysis to introduce the electric current
averaged with respect to the membrane length $l$ and with respect to the time:
\begin{equation}\label{eq76}
\langle j(t) \rangle = \frac{1}{l}\int_0^l j(x,t)\,dx,
 \quad J =\lim_{T \to \infty}\frac{1}{T}\int_0^T \langle j(t)\rangle\,dt.
\end{equation}

The problem is characterized by three dimensionless parameters: $\Delta V$,
$\nu$, which is a small parameter, and~$\varkappa$. The dependence on the
concentration, $p$, for the overlimiting regimes is practically absent and thus
$p$ is not included in the mentioned parameters.

The problem was solved for $\nu=10^{-4}\div 10^{-3}$, $\varkappa=0.02\div 0.5$,
and the dimensionless potential drop varied within $\Delta V=0\div100$. In all
calculations, $p=5$. (Some of the calculations were executed for $p=1$ and
$p=2$. For the overlimiting regimes, the results coincide with a graphical
accuracy with the ones for $p=5$.)

Just to provide perspective, the bulk concentration of the aqueous electrolytes
varies in the range $\tilde{c}_{\infty}=1 \div 10^3\text{ mol}/\text{m}^3$; the
potential drop is about $\Delta\tilde{V}=0 \div 5$~V; the absolute temperature
can be taken as $\tilde{T}=300$~K; the diffusivity is about
$\tilde{D}=2\cdot 10^{-9} \div 10^{-8}\text{ m}^2/\text{s}$; the distance
between the electrodes $\tilde{L}$ is of the order of $0.5 \div 1.5$~mm; the
concentration value $\tilde{p}$ on the membrane surface must be much larger than
$\tilde{c}_{\infty}$ and it is usually taken within the range from
$5\tilde{c}_{\infty}$ to $10\tilde{c}_{\infty}$. The dimensional Debye layer
thickness $\tilde{\lambda}_D$ varies in the range from 1 to 100~nm, depending on
the concentration $\tilde{c}_{\infty}$.

\section{The numerical method}

\subsection{Discretization of the \NPPS\ system}
A~finite-difference method with second-order accuracy is used for the spatial
discretization of \eqref{eq1}--\eqref{eq3}. In accordance with a standard
staggered grid layout, all scalars ($c^{\pm}, \rho, \Phi, \Pi$) are defined in
the centers of the computational cells, while the velocity components are sought
in the centers of the cell faces. The grid is stretched in the normal
$y$\nobreakdash-direction via a $\tanh$ stretching function in order to properly
resolve the thin boundary layers attached to the membrane surfaces. A~uniform
grid is used in the homogeneous tangential $x$\nobreakdash-direction. The
discretization of the linear terms in \eqref{eq1}--\eqref{eq3} as well as the
formulation of the boundary conditions are straightforward. The concrete
expressions can be found elsewhere, in \cite{Niki2} for example. The nonlinear
terms in \eqref{eq1}, written in a divergent form
$\vec{u}\cdot\nabla c^\pm\mp\nabla\cdot\left(c^\pm\nabla\Phi\right)\equiv\nabla\cdot\left(c^\pm(\vec{u}\mp\nabla\Phi)\right)$, are discretized by
central differences with the use of a second-order interpolation of $c^\pm$ from
the cell centers to the cell faces.

As a result of the spatial discretization, the problem is reduced to the
solution of the set of ordinary differential equations:
\begin{equation}
\frac{dC}{dt}=F(C). \label{eq4}
\end{equation}
Here, the $2N$\nobreakdash-dimensional vector $C(t)$ is composed of $N$ grid
values of $c^+$ and $N$ grid values of $c^-$, where $N=N_xN_y$ is the total
number of cells in the computational mesh and $N_x$ and $N_y$ are the numbers of
grid points in the respective spatial directions. The vector function $F(C)$ in
\eqref{eq4} includes the discretized nonlinear and linear terms of \eqref{eq1}.

The grid functions $\Phi$ and $\vec{u}$ which are required for the calculation
of $F$ are found for a given $C$ from the discretized Poisson equation
\eqref{eq2} and discretized Stokes problem \eqref{eq3}. These linear problems
are solved using the fast Fourier transformation (FFT) in the homogeneous
direction with subsequent solution of the resulting one-dimensional equations.
Thus, the discretized Poisson and Stokes problems are solved exactly by direct
methods without iteration.

Equation \eqref{eq4} represents a stiff problem when a fine spatial resolution
is used. Explicit methods of time advancement are prohibitively ineffective for
this problem. Fully implicit methods require the inversion of the nonlinear
right-hand side operator $F(C)$ and thus are extremely costly. A~reasonable
compromise may be a semi-implicit method, where only a part (usually linear) of
the operator $F(C)$ is treated implicitly. To clarify the idea of a
semi-implicit method as applied to \eqref{eq4}, let us consider the first-order
accurate semi-implicit Euler method. The advancement from time instant $t_0$ to
the time instant $t_1=t_0+\Delta t$ in this method is made according to the
following equation:
\begin{equation}
\frac{C_1-C_0}{\Delta t}=F(C_0)-LC_0+LC_1. \label{eq5c}
\end{equation}
Here $C_0=C(t_0)$ is the given vector and $C_1=C(t_1)$ is to be found. The
linear operator $LC$ is extracted from $F(C)$ and treated implicitly in order to
make time integration procedure more stable. It is appropriate to rewrite
\eqref{eq5c} in the following equivalent form:
\begin{equation}
(I-\Delta tL)\Delta C = \Delta tF(C_0), \quad \Delta C=C_1-C_0 \label{eq6}
\end{equation}
with $I$ as the identity operator: $IC\equiv C$. One can see that the scheme
\eqref{eq6} possesses first-order accuracy for any operator~$L$. Thus the
concrete form of the implicit operator $L$ does not affect the formal accuracy
of the scheme, but it does affect its stability. A~proper choice of the implicit
operator determines the performance of the method. One extreme form of $L$ is
the zero operator $L\Delta C\equiv0$ which makes the scheme \eqref{eq6}
explicit. The opposite extreme is $L=J$, where $J=dF(C)/dC$ is the Jacobian
matrix of the problem. Probably, $L=J$ ensures the most stable time integration
and thus allows the largest possible value of the time step $\Delta t$. However,
the solution of a linear algebraic system with a full matrix $(I-\Delta tL)$,
which is what is involved in \eqref{eq6}, becomes prohibitively costly. So, a
compromise must be found where $L$ is reasonably simple (say, sparse or banded),
so that \eqref{eq6} can be solved efficiently and at the same time $L$ must be
in a some sense close to the Jacobian~$J$. Closeness to the Jacobian means that
$L$ must contain those terms of $J$ that are responsible for the stiffness of
the problem.

In the present paper, the semi-implicit Runge-Kutta scheme with third-order
accuracy from \cite{Niki1} is used for the time integration of \eqref{eq4}. At
each step of the time advancement, this scheme requires a triple calculation of
the right-hand side function $F(C)$ and four solutions of a linear problem
similar to \eqref{eq6}. The scheme is equipped with an algorithm of local
accuracy estimation and automatic time-step control. Such an option is
especially useful when searching for an optimal form of implicit operator~$L$.
Since the minimum mesh size in the direction normal to the wall is typically two
or even three orders of magnitude smaller than those in the tangential
directions, the largest part of the stiffness of the problem is connected with
the second derivatives in this direction, namely, with the terms approximating
$\partial^2 c^\pm/\partial y^2$ in the right-hand side of \eqref{eq1}. If only
these terms are treated implicitly, the \eqref{eq6} takes the form
\begin{equation}
\frac{\delta^2\Delta C}{\delta y^2}-\frac{1}{\Delta t}\Delta C=-F(C_0)
\label{eq7}
\end{equation}
where $\delta/\delta y$ stands for the finite-difference derivative in the
$y$\nobreakdash-direction. Eq.~\eqref{eq7} presents a system with a
three-diagonal matrix and is solved by the standard sweep method. The use of an
implicit operator of the form \eqref{eq7} liberalizes the stability restrictions
appreciably and raises the maximal permissible value of $\Delta t$ by a factor
of $10^3 \div 10^4$ over that of an explicit time advancement scheme at typical
values of the problem and grid parameters used in the simulations.

Although a limited amount of simulation can be performed (and actually was
performed) with the use of \eqref{eq7}, the stability restrictions are still
evident. By modifying the implicit operator it was found that the second
important source of instability comes from the term approximating the
wall-normal counterpart of the term $\pm\nabla\cdot(c^\pm\nabla\Phi)$ in the
right-hand side of \eqref{eq1}, namely,
$\pm\delta(c^\pm\delta\Phi/\delta y)/\delta y$ (with partial derivatives
$\partial/\partial y$ replaced with their finite-difference analogs
$\delta/\delta y$). Taking in the last expression $c^\pm=c^\pm_0+\Delta c^\pm$
and linearizing near $c^\pm_0=c^\pm(t_0)$ yields
$\pm\delta(c^\pm_0\delta\Delta\Phi/\delta y +
\Delta c^\pm\delta\Phi_0/\delta y)/\delta y$. Here $\Phi_0$ and $\Delta\Phi$ are
linearly expressed in terms of $c^\pm_0$ and $\Delta c^\pm$, respectively,
through the finite-difference analog of the Poisson equation \eqref{eq2}. One
more simplification which makes it feasible to add the last expression into the
implicit operator is to neglect the derivatives along the tangential direction
in the Poisson equation for $\Delta\Phi$. Thus, \eqref{eq6} with the modified
implicit operator becomes the linear problem
\begin{gather}
\frac{\delta^2\Delta c^\pm}{\delta y^2}
 \pm\delta(c^\pm_0\delta\Delta\Phi/\delta y
 + \Delta c^\pm\delta\Phi_0/\delta y)/\delta y - \frac{1}{\Delta t}\Delta c^\pm
 = -F^\pm(C_0); \label{eq8}\\
\nu^2\frac{\delta^2\Delta\Phi}{\delta y^2} = \Delta c^--\Delta c^+. \label{eq9}
\end{gather}
The linear system \eqref{eq8}--\eqref{eq9} has a sparse matrix and is solved
with the use of the IMSL routine \texttt{LSLXG}. The use of
\eqref{eq8}--\eqref{eq9} instead of \eqref{eq7} produces a further increase of
$\Delta t$ by two orders of magnitude. Although solving \eqref{eq8}--\eqref{eq9}
via the \texttt{LSLXG} routine is up to 10 times longer than solving
\eqref{eq7}, the overall advantage of using \eqref{eq8}--\eqref{eq9} instead of
\eqref{eq7} is obvious.

\subsection{Domain length and discretization of the initial conditions}
For the natural ``room disturbances'' the infinite spatial domain was changed to
the large enough finite domain that has length $l$, with the corresponding wave
number $k=2\pi/l$. The condition that the solution should be bounded as
$x \to \infty$ was changed to periodic boundary conditions. The length of the
domain $l$ had to be taken large enough to make the solution independent of the
domain size. The wave number $k$ was taken to be 1, while $k=0.5$ and $k=0.25$
were used to verify the results.

The initial conditions \eqref{eqq67} in discrete form are
\begin{gather}
t=0: \nonumber\\
c^+ -1 \approx \sum_{m=1}^{M}|\hat{p}(k_m)| \exp{(-i k_m x +i\theta_m^{+})},
\quad c^- -1 \approx \sum_{m=1}^{M}|\hat{n}(k_m)| \exp{(-i k_m x
+i\theta_m^{-})}, \label{x1}
\end{gather}
where the integrals were approximated by the sum, $k_m = m \cdot k$, $k=k_c/M$,
$k_c$ is the largest wave number taken into account, and $\theta_m^{\pm}$ are
the phases of the complex amplitudes $\hat{p}$ and $\hat{n}$, which were set by
the random number generator with a uniform distribution in the interval
$[0,2\pi]$. It was assumed that the amplitudes $|\hat{p}|$ and $|\hat{n}|$ are
zero above $k_c$ and equal to a certain small value describing the natural
external noise inside the interval $0 < k_m < k_c$. To mimic a broadband white
noise in our calculations $|\hat{p}|=|\hat{n}|=10^{-6}$ were taken, but some of
the calculations were carried out with initial perturbations 100 times larger:
$|\hat{p}|=|\hat{n}|=10^{-3}$. All the calculations were limited by the final
time $t_{\max}=6 \div 45$, which corresponds to the dimensional time from
$\tilde{t}_{\max}$ of 5 to 60 minutes. In most of the presented calculations,
$\nu=10^{-3}$ was taken.

\section{Simulation results}

In this section the results of the simulation will be analyzed using the
standard tools of the theories of hydrodynamic stability and nonlinear dynamical
systems. Unless otherwise stated, $\nu=10^{-3}$ or $\nu=10^{-4}$.

\subsection{1D trivial solution and its stability}
\begin{figure}[p]
\centering
\includegraphics[width=\textwidth]{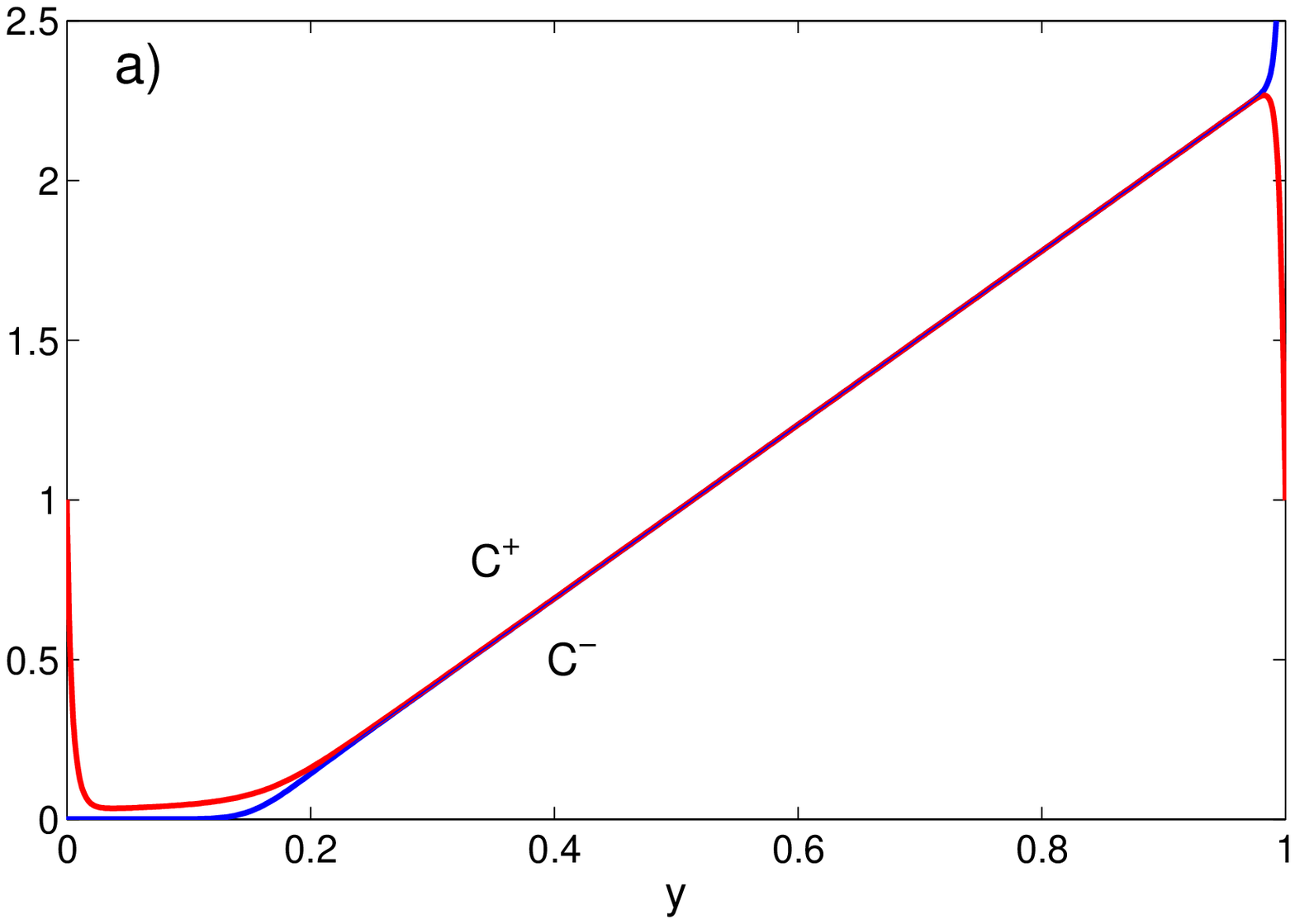}
\includegraphics[width=\textwidth]{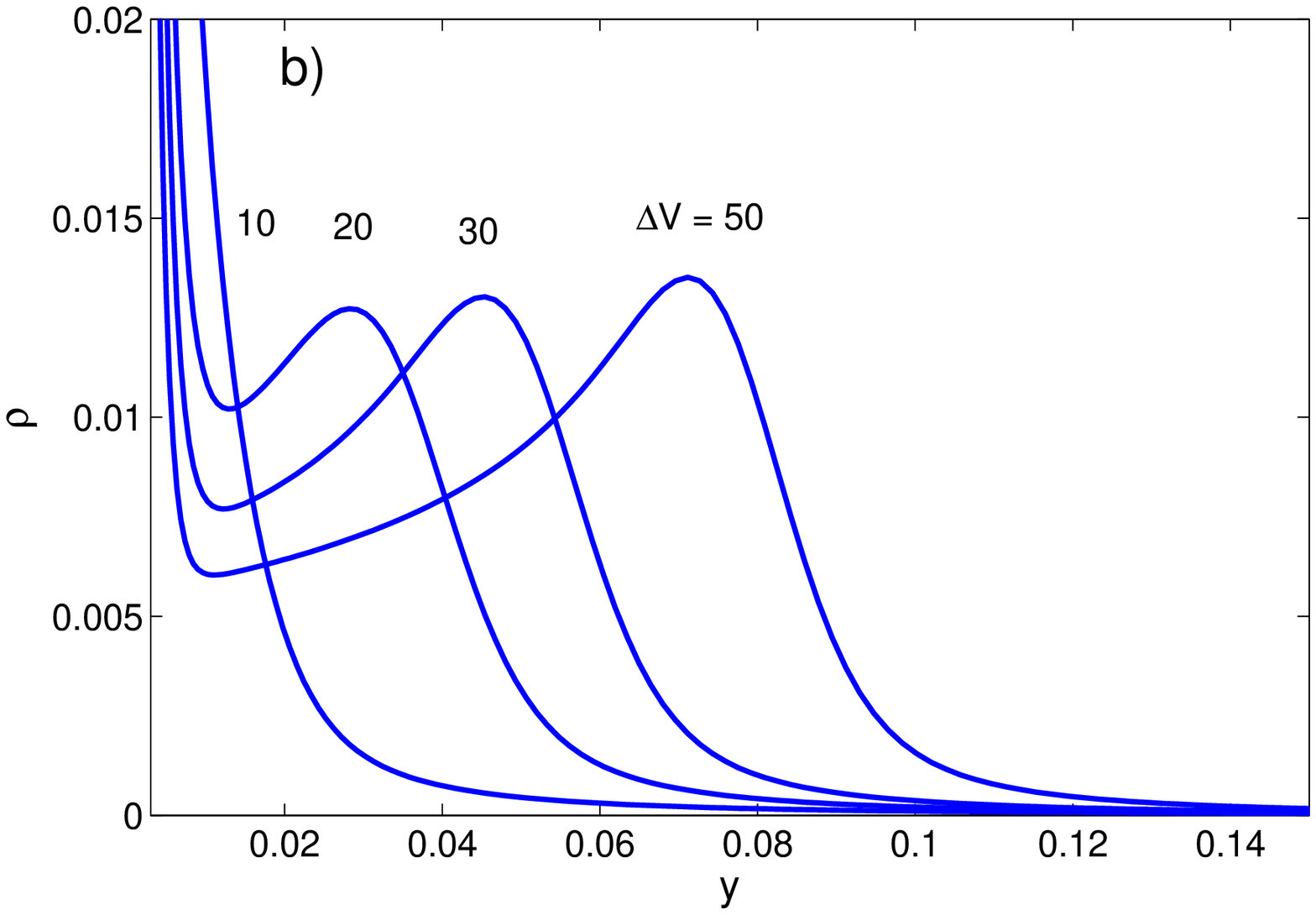}
\caption{(Color online) 1D steady solution. Ion concentration for $\Delta
V=25$, $\varkappa=0.1$ and $\nu=7\times 10^{-3}$ (a); charge density (b) near
the membrane for $\nu=10^{-3}$, $\varkappa=0.1$ and different $\Delta V$}
\label{stat}
\end{figure}

The simulations showed that for $\Delta V < \Delta V_* \approx 29.5$
one-dimensional steady solutions were realized. For
$\partial/\partial t=\partial/\partial x=0$ the system
\eqref{eq1}--\eqref{eq3} turns into
\begin{equation}\label{L1}
c^+\frac{d\Phi}{dy} + \frac{dc^+}{dy} = 4j, \quad c^-\frac{d\Phi}{dy}
 + \frac{dc^-}{dy} = 0, \quad \nu^2\frac{d^2\Phi}{dy^2} = c^- - c^+
\end{equation}
with the boundary conditions
\begin{equation}\label{L2}
y=0: \quad c^+=p, \quad \Phi=0,
 \qquad y=1: \quad c^+=p, \quad \Phi=\Delta V, \qquad \int_0^1 c^{-} dy=1.
\end{equation}
This system describes the 1D solution, which is decoupled from the
hydrodynamics,  $u=v=0$. The first numerical solution was obtained in
\cite{Rub1} for the asymptotical solution, $\nu \to 0$, see
\cite{Bab5,Chu1,Rub12}. Fig.~\ref{stat} shows typical profiles of $c^{\pm}$ and
$\rho$ for the underlimiting and limiting regimes.
Numerical simulation of the full system \eqref{eq1}--\eqref{eqq6} is in good
agreement with the results obtained in \cite{Rub1}.

At $\Delta V=\Delta V_*$ the 1D solution \eqref{L1}--\eqref{L2} becomes unstable
with respect to sinusoidal perturbations with wave number~$k$,
\begin{equation}\label{L2a}
c^{\pm}=c_0^{\pm}+\hat{c}^{\pm}\exp{(ikx+\lambda t)},
 \quad \Phi=\Phi_0+\hat{\Phi}\exp{(i k x+\lambda t)},
 \quad \Psi=\hat{\Psi} \exp{(ikx+\lambda t)}.
\end{equation}
Suppose that these perturbations trigger a hydrodynamic flow, so that now
$\Psi \ne 0$. The subscript $0$ is related to the mean 1D solution; hat, to
perturbations. Upon linearization of \eqref{eq1}--\eqref{eq5} with respect to
perturbations and omitting the subscript $0$ in the mean solution, we get the
system
\begin{equation}\label{L3}
\lambda \hat{c}^{+}-i k \frac{dc^{+}}{dy}\hat{\Psi} =\frac{d}{d y}\left(
c^{+}\frac{d \hat{\Phi}}{dy}+\frac{d \Phi}{dy}\hat{c}^{+}+\frac{d
\hat{c}^{+}}{dy} \right)-k^2 c^+ \hat{\Phi}-k^2\hat{c}^{+},
\end{equation}
\begin{equation}\label{L4}
\lambda \hat{c}^{-}-i k \frac{d c^{-}}{dy}\hat{\Psi} =\frac{d}{d y}\left(-
c^-\frac{d \hat{\Phi}}{dy}-\frac{d \Phi}{dy}\hat{c}^{-}+\frac{d
\hat{c}^{-}}{dy} \right)-k^2 c^- \hat{\Phi}-k^2\hat{c}^{-},
\end{equation}
\begin{equation}\label{L5}
\nu^2 \left( \frac{d^2\hat{\Phi}}{dy^2} -k^2 \hat{\Phi }\right)
=\hat{c^{-}}-\hat{c^{+}} \equiv -\hat{\rho},
\end{equation}
\begin{equation}\label{L6}
\frac{d^4 \hat{\Psi}}{dy^4}-2 k^2 \frac{d^2\hat{\Psi}}{dy^2} +k^4 \hat{\Psi}= i
k \frac{\varkappa}{\nu^2}\left(\frac{d\rho}{dy}\hat{\Phi}-\frac{d
\Phi}{dy}\hat{\rho} \right),
\end{equation}
\begin{equation}\label{eq70}
y=0: \quad \hat{\Phi}=0, \quad \hat{\Psi}=0,\quad \:\frac{d \hat{\Psi}}{dy}=0,
 \quad c^-\frac{d\hat{\Phi}}{dy}+\frac{d \Phi}{dy}\hat{c}^{-}-\frac{d \hat{c}^{-}}{dy}=0, \quad \hat{c}^{+}=0
\end{equation}
\begin{equation}\label{eq71}
y=1: \quad \hat{\Phi}=0, \quad \hat{\Psi}=\frac{d \hat{\Psi}}{dy}=0,\quad
c^-\frac{d \hat{\Phi}}{dy}+\frac{d\Phi}{dy}\hat{c}^{-}-\frac{d
\hat{c}^{-}}{dy}=0 ,\quad \hat{c}^+=0,
\end{equation}
which is an eigenvalue problem for~$\lambda$. If $\lambda(k)<0$ for all wave
numbers $k$, the 1D solution \eqref{L1}--\eqref{L2} is stable; if $\lambda(k)>0$
for at least one~$k$, the 1D solution is unstable and it gives the first
bifurcation point of the system. The discretization was realized by expanding
the unknown functions in Chebyshev series plus using the Lanczos procedure to
satisfy the boundary conditions. The matrix eigenvalues were calculated by the
QR-algorithm.

\begin{figure}[hbtp]
\centering
\includegraphics[angle=-90,width=\textwidth]{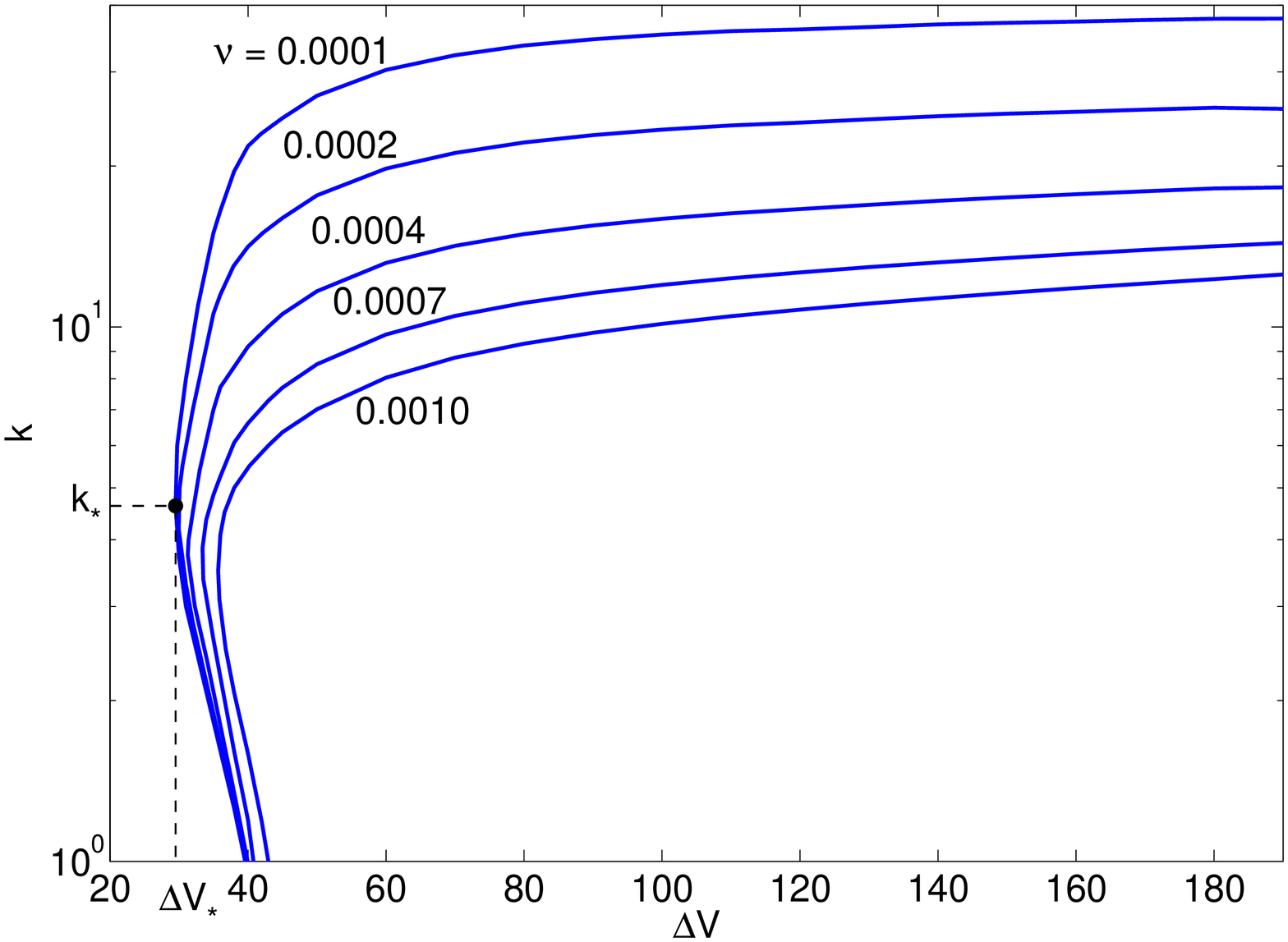}\\
\begin{picture}(0,0)
\put(-60,40){\includegraphics[width=10cm]{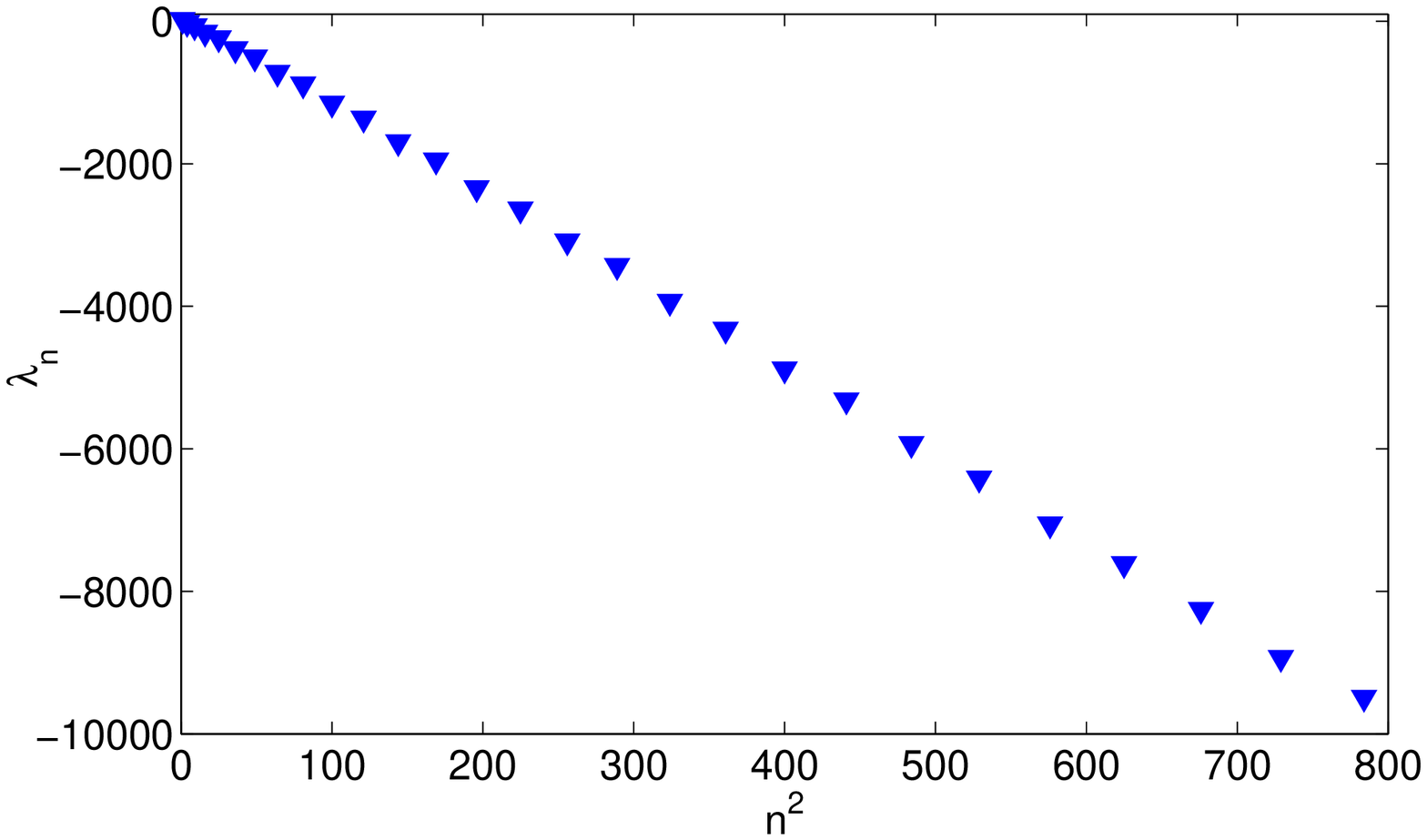}}
\end{picture}
\caption{(Color online) Marginal stability curves $k$ vs $\Delta V$ for
$\varkappa=0.1$ and different values of the parameter $\nu$. Inset: discrete
spectrum $\{\lambda_k\}$ of the stability problem \eqref{L3}--\eqref{eq71} for
$k=4$, $\Delta V=29$, $\varkappa=0.1$ and $\nu=10^{-3}$}
\label{Lk01}
\end{figure}

In all our calculations the $\lambda_k$ were real numbers, which means that the
instability is monotonic. These eigenvalues are arranged in such a way that
$\lambda_1 > \lambda_2 > \lambda_3 > \dots$. The inset of Fig.~\ref{Lk01} shows
that for large $k$, $\lambda_k \sim k^2$. Fig.~\ref{Lk01} pictures the marginal
stability curves, $\lambda_1(k)=0$, in the plane $\Delta V - k$ for
$\varkappa=0.1$ and various values of the parameter~$\nu$.

The results of the numerical simulation of the nonlinear system
\eqref{eq1}--\eqref{eqq6} with the natural white noise initial conditions
\eqref{eqq67} are in good correspondence with the linear stability results which
are presented in Table~\ref{T3}.
\begin{table}[hbt]
\centering
\caption{Critical parameters for $\nu=10^{-3}$ and different $\varkappa$}
\label{T3}
\begin{tabular}{|c|c|c|c|c|c|c|}
\hline
$\varkappa$ & 0.02 & 0.05 & 0.10 & 0.15 & 0.20 & 0.50\\
\hline
$\Delta V_*$& 55.0 & 38.0 & 29.5 & 25.5 & 23.5 & 19.1\\
\hline
$k_*$ & 4.65 & 4.7 & 4.8 & 4.9 & 5.0 & 5.1\\
\hline
\end{tabular}
\end{table}
Note that the numerical approach of \cite{Dem} and the asymptotical approach
\cite{Rub13} for small $\nu$ are also in good correspondence with the numerical
simulation of the nonlinear system (our Debye number $\nu$ and the $\varepsilon$
used in \cite{Rub13} are related by $\varepsilon=\nu/\sqrt{2}$; the
$Pe$\nobreakdash-number of \cite{Rub13} is identical to our $\varkappa$). The
transition to the primary instability to the critical point was approached both
from below and from above in $\Delta V$, and no hysteresis was observed,
indicating that the bifurcation is supercritical for the parameter values
studied here.

\begin{figure}[hbtp]
\centering
\includegraphics[width=\textwidth]{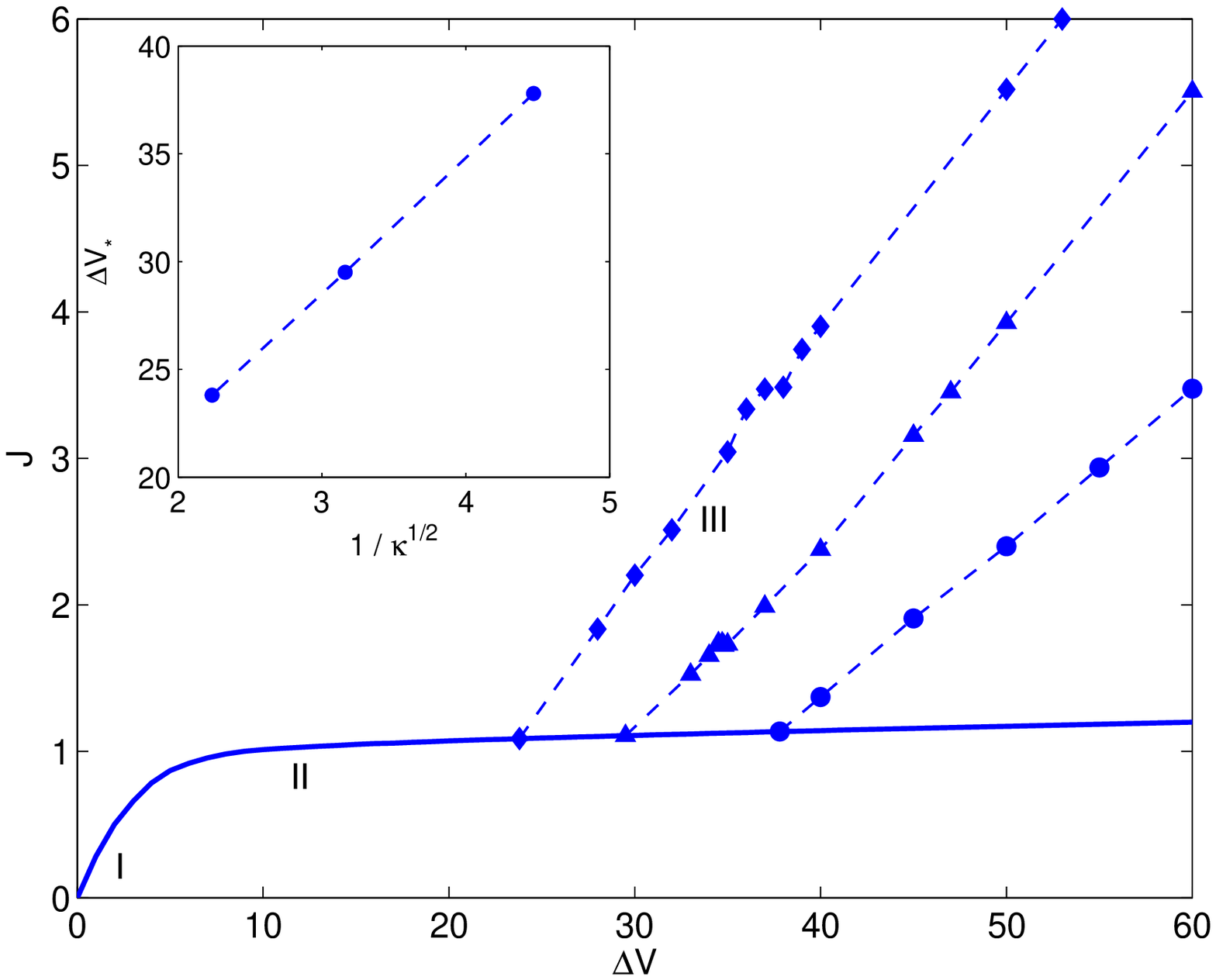}
\caption{(Color online) VC characteristics for $\varkappa=0.2$ (diamonds),
$0.1$ (triangles) and $0.05$ (circles) and $\nu = 10^{-3}$. I, II, and III are
the underlimiting, limiting, and overlimiting currents, respectively. The inset
shows that $\Delta V_*$ linearly depends on $1/\sqrt{\varkappa}$; as $\varkappa
\to \infty$, $\Delta V_* \to 8.94$}
\label{aaa10}
\end{figure}

The critical potential drop $\Delta V_*$ and critical wave number $k_*$ for
different $\varkappa$ are presented in Table~\ref{T3}. The instability triggers
an additional mechanism of ion transport by convection and, hence, it is
reasonable to visualize this bifurcation in the volt--current (VC)
characteristic, see Fig.~\ref{aaa10}. The VC characteristics for the
underlimiting and limiting regimes (solid curve in the figure) are described by
the system \eqref{L1}--\eqref{L2} and do not depend on the coupling coefficient
$\varkappa$, because no hydrodynamics is involved in the process. The curves
corresponding to the overlimiting regime and different $\varkappa$ depart from
this solid line at different bifurcation points $\Delta V*$, depending
on~$\varkappa$. The smaller is $\varkappa$, the more distant is the branching
point from the origin. The critical potential drop $\Delta V_*$ versus the
coupling coefficient $\varkappa$ (see the inset to Fig.~\ref{aaa10}) shows that
$\Delta V_* = \const_1+\const_2/\sqrt{\varkappa}$ can be used as a good
interpolation relation. Increasing the coupling coefficient $\varkappa$, the
plateau region of the limiting regimes decreases, and in the limit
$\varkappa \to \infty$, $\Delta V_* \to 8.95$. Note that the last value
approximately corresponds to the potential drop of the transition from
underlimiting regimes to limiting regimes. The dependences for the critical drop
of potential and wave number can be presented in the following form,
\[
\nu = 10^{-3}, \quad \Delta V_* = 8.95 + \frac{6.52}{\sqrt{\varkappa}},
 \quad k_*-4.8 = 2(\varkappa-0.1).
\]

The dependence of $k_*$ on $\varkappa$ is rather weak and, very roughly,
$k_*\approx 4.8$ for all considered~$\varkappa$. In order to get $k_*$
accurately, the spatial interval was doubled and even tripled.

\begin{table}[hbt]
\centering
\caption{Bifurcation points, $\varkappa=0.1$}\label{T2}
$\vphantom{\Biggl(}
\begin{array}{|c|c|c|c|}
\hline
\Delta V_* & \Delta V_{**} & \Delta V_{***} \\
\hline
29.5 & 32.5 & 37.0 \\
\hline
\end{array}$
\end{table}

\subsection{Weakly nonlinear analysis and behavior near threshold}
For the control parameter $\Delta V > \Delta V_*$ the uniform 1D state suffers a
supercritical bifurcation. At the threshold of instability $k_* \ne 0$ and
according to the classification of \cite{CrHog} the instability is spatially
periodic. Near $\Delta V_*$, $k_*$ the broadband initial spectrum \eqref{x1} is
filtered by the instability into a narrow band near the wave number $k_*$ and
has a sinusoidal form with a slow envelope,
\[ a\cdot e^{ik_*x}. \]

\begin{figure}[hbtp]
\centering
 \includegraphics[width=\textwidth]{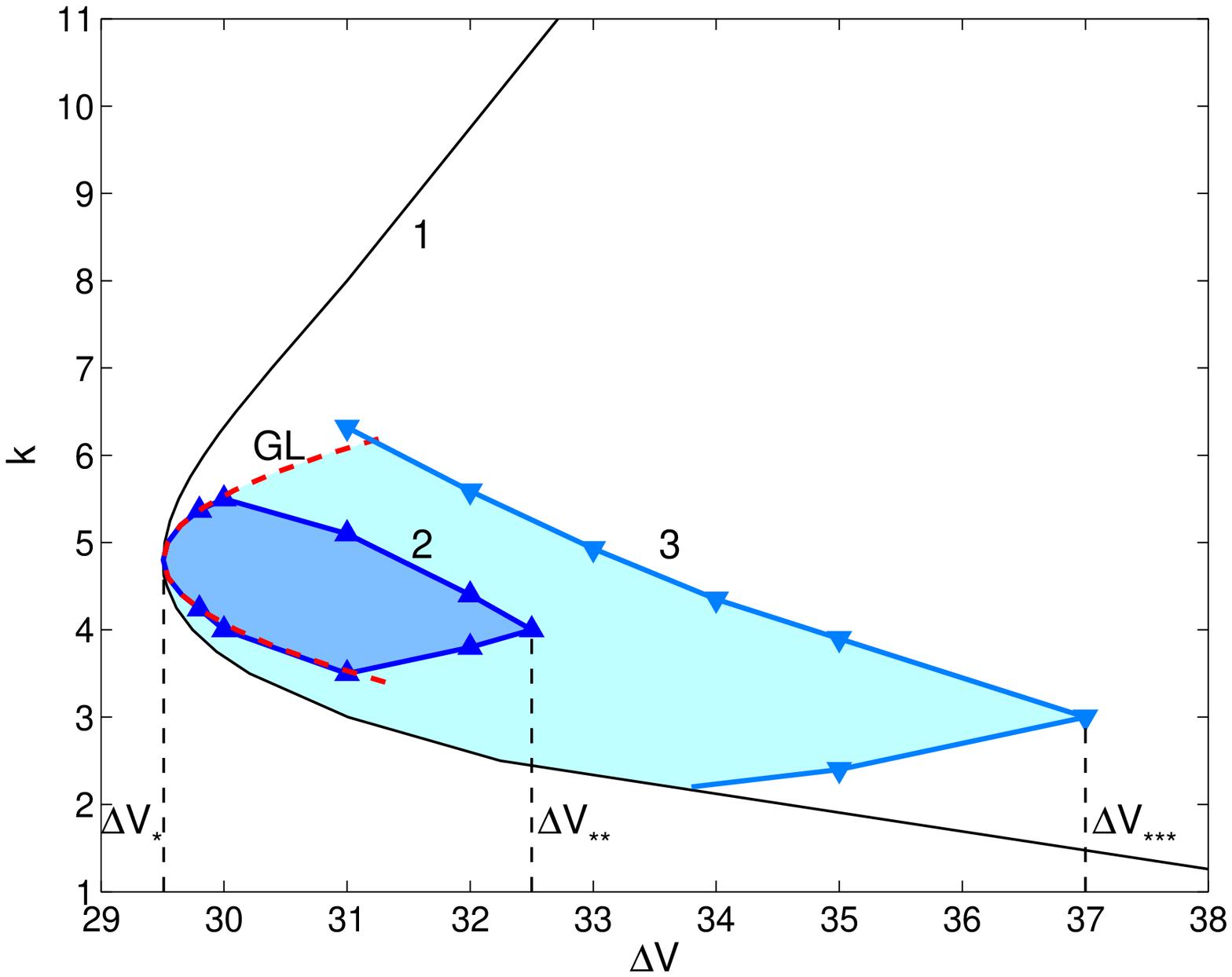}
\caption{(Color online) Inside the curve 1 the trivial 1D solution is unstable;
inside the region bounded by 2 there is a wave-number window of stable 2D
steady solutions, this window according to GL-equation is marked in the figure
as GL; for the region inside 3 the saddle value $S$ is negative. Triangles are
taken from our simulation, $\varkappa=0.1$}
\label{egg}
\end{figure}

For our kind of instability, near the point $\Delta V=\Delta V_*$ and $k=k_*$,
the nonlinearities of the entire system \eqref{eq1}--\eqref{eq7} are weak and
the spatial and temporal modulations of the basic linear stability pattern
$ae^{ik_*x}$ are slow and their narrow band near $k_*$ is described by the
weakly nonlinear Ginzburg-Landau (GL) equation (see \citetwo{CrHog}{WE}),
\begin{equation}\label{GL1}
\frac{\partial a}{\partial t}=\gamma_1(\Delta V-\Delta V_*)\cdot a+\gamma_2
\frac{\partial^2 a}{\partial x^2}-\gamma_3|a|^2a + \text{higher order terms}.
\end{equation}
The first term in the right-hand side of the equation characterizes the
instability and $\gamma_1$ is a real positive number. The second and third terms
are responsible, respectively, for the dissipation and the nonlinear saturation,
$\gamma_2$ and $\gamma_3$ for the common case are complex numbers, but in our
case they are real numbers; these coefficients must be calculated numerically.
Eq.~\eqref{GL1} is valid for a small supercriticality,
$\Delta V - \Delta V_* = \delta \to 0$. Upon rescaling the amplitude, time, and
coordinate,
\[
a = \left(\frac{\gamma_1}{\gamma_3}\right)^{1/2}\delta^{1/2} A,
 \quad\frac{\partial}{\partial t} = \gamma_1\delta\frac{\partial}{\partial\tau},
 \quad \frac{\partial^2}{\partial x^2}
 = \left(\frac{\gamma_1}{\gamma_2}\right)\delta\frac{\partial^2}{\partial\xi^2},
\]
eq.~\eqref{GL1} turns into
\begin{equation}\label{GL2}
\frac{\partial A}{\partial\tau} = A+\frac{\partial^2 A}{\partial\xi^2}-|A|^2A.
\end{equation}
The 1D solution \eqref{L1}--\eqref{L2} corresponds to the trivial solution of
\eqref{GL2}, $A=0$. Eq.~\eqref{GL2} also has the one-parameter family of
nontrivial solutions,
\begin{equation}\label{GL3}
A = \sqrt{1-\alpha^2} \cdot \exp({i\alpha\xi}),
\end{equation}
with the parameter $\alpha$, $-1<\alpha<1$. This parameter is connected with the
wave number $k$ by the relation
\begin{equation}\label{GL3a}
\alpha = \left(\frac{\gamma_2}{\gamma_1}\right)^{1/2}(k-k_*)/\delta^{1/2}.
\end{equation}
The solutions \eqref{GL3} are stable within the window
$-\sqrt{3}/3<\alpha<\sqrt{3}/3$ (see \cite{CrHog}). Taking into account
\eqref{GL3a}, this window can be recalculated to the stability region for the
original wave number~$k$,
\begin{equation}\label{GL3b}
k_* - \sqrt{\frac{\delta\gamma_1}{3\gamma_2}} < k < k_*
 + \sqrt{\frac{\delta\gamma_1}{3\gamma_2}}.
\end{equation}
This stability domain is denoted by GL in Fig.~\ref{egg}.

\subsection{Family of 2D steady space-periodic solutions and their stability}
\begin{figure}[p]
\centering\begin{tabular}{@{}r@{}}
\includegraphics[width=.6\textwidth]{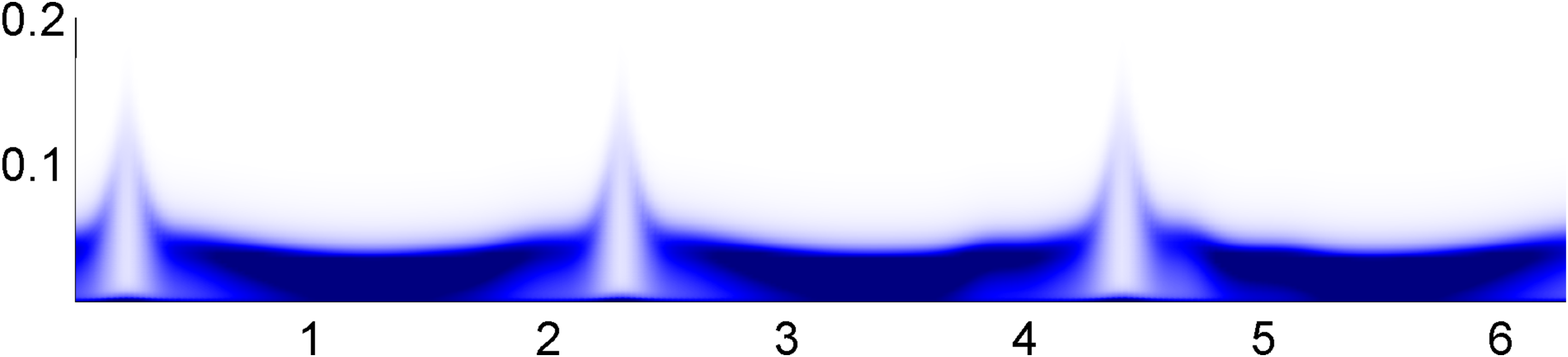}\raisebox{4.5em}[0em][0em]{\makebox[0em]{\hspace{-1em}(a)}}\\
\includegraphics[width=.62\textwidth]{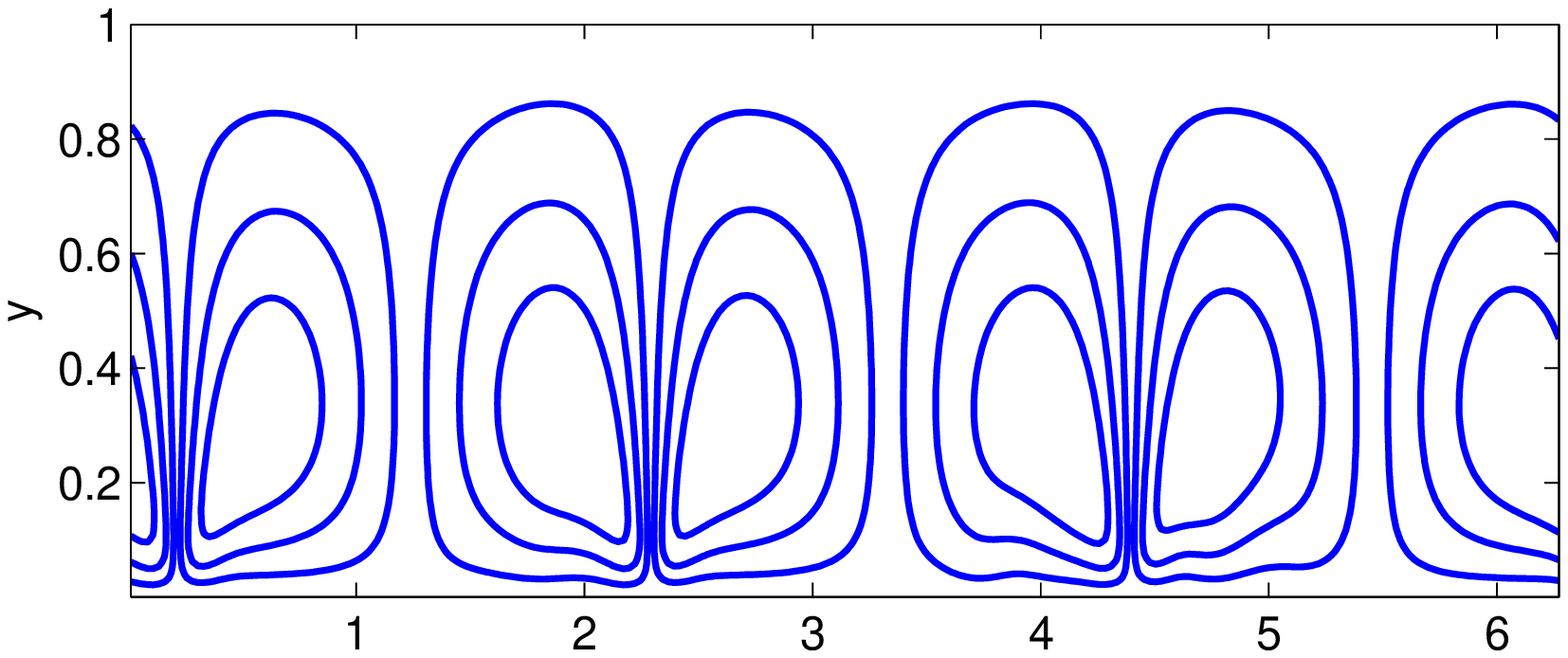}\\
\includegraphics[width=.6\textwidth]{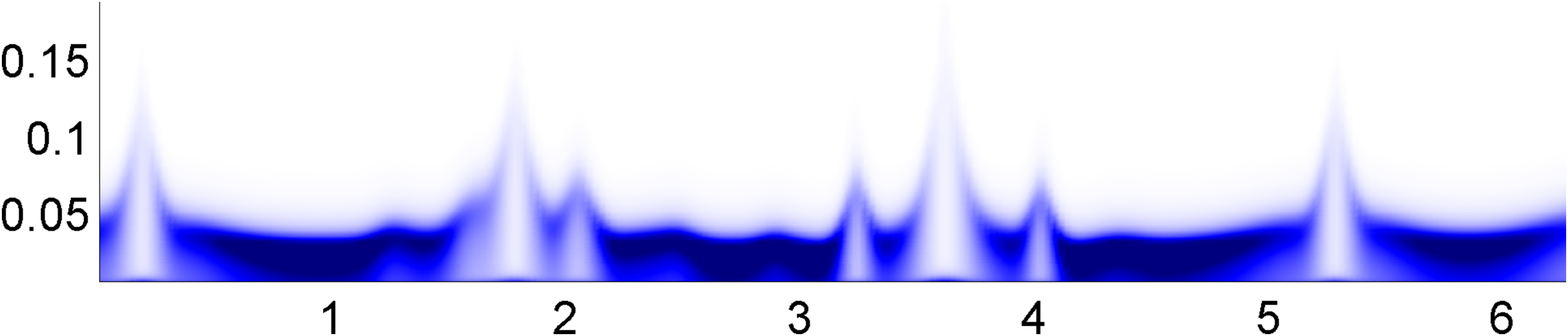}\raisebox{4.5em}[0em][0em]{\makebox[0em]{\hspace{-1em}(b)}}\\
\includegraphics[width=.62\textwidth]{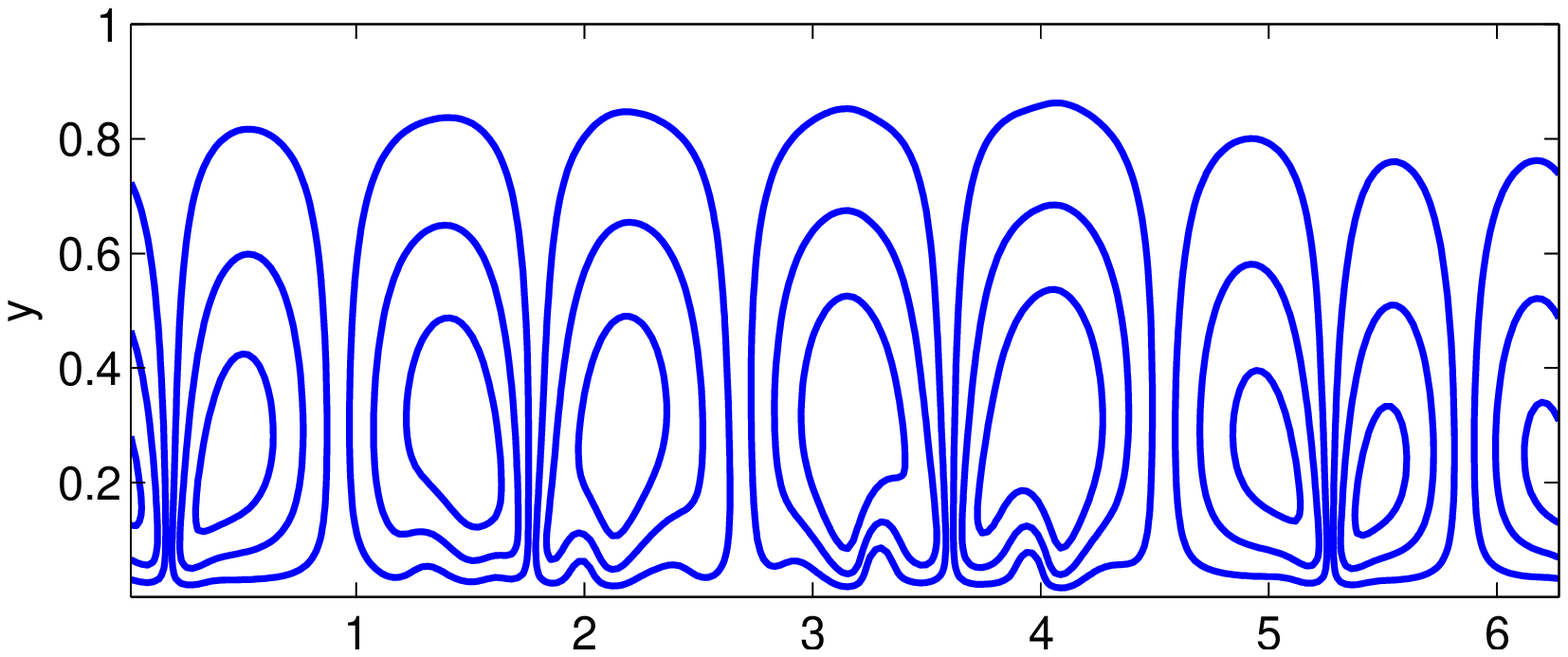}\\
\includegraphics[width=.6\textwidth]{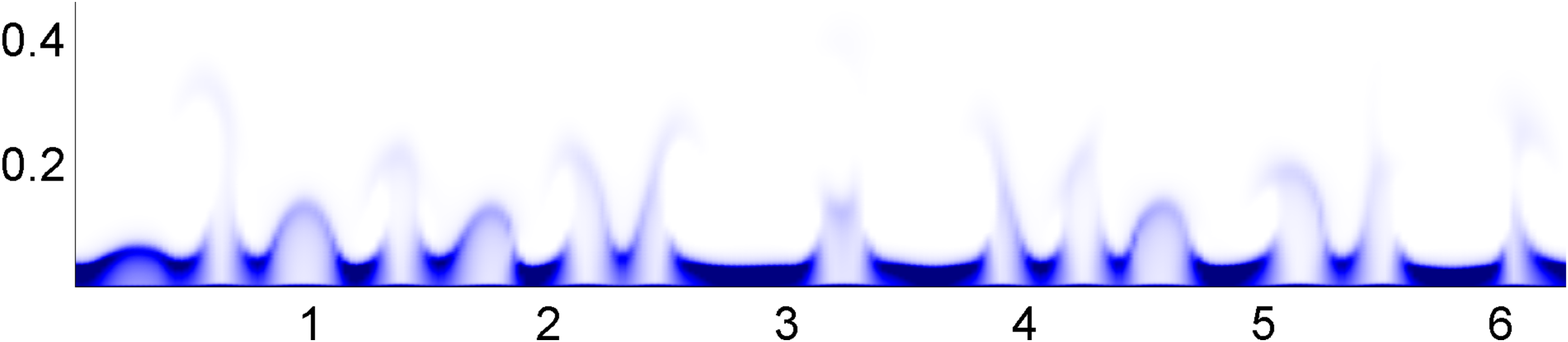}\raisebox{4.5em}[0em][0em]{\makebox[0em]{\hspace{-1em}(c)}}\\
\includegraphics[width=.62\textwidth]{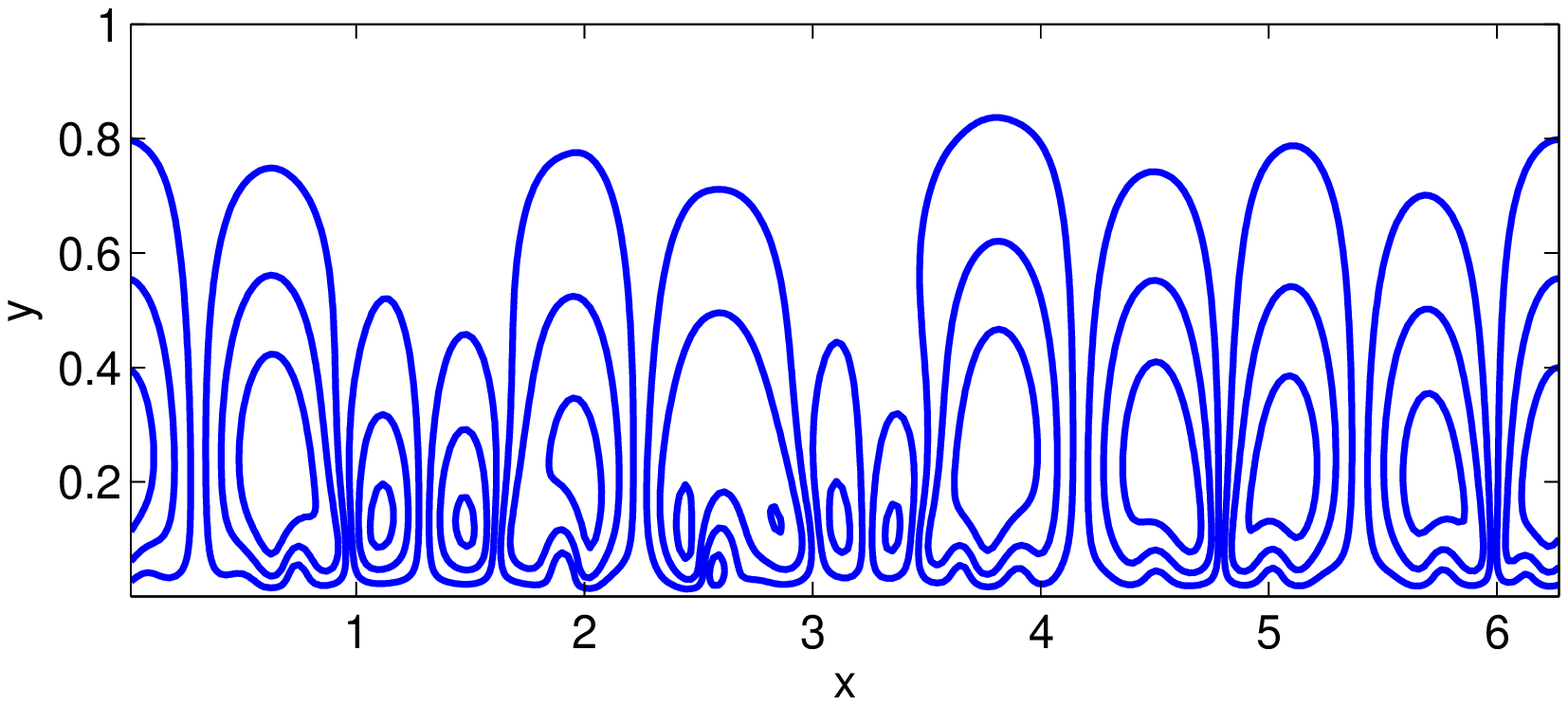}
\end{tabular}
\caption{(Color online) Typical snapshots of the charge density $\rho(x,y,t)$
and the stream-lines $\psi(x,y,t)$ at the end of evolution; darker regions
correspond to larger charge densities. (a) $\Delta V=31.5$, (b) $\Delta V=35$,
(c) $\Delta V=50$, $\varkappa=0.1$ and $\nu=10^{-3}$}
\label{aaa12}
\end{figure}

Numerical simulations of the entire system \eqref{eq1}--\eqref{eq5} with the
natural white noise initial conditions \eqref{eqq67} confirmed the
aforementioned predictions of the weakly nonlinear analysis: in some vicinity of
the critical point, nonlinear saturation led to steady,
$\partial/\partial t=0$, space-periodic 2D solutions. We call them steady
periodical electroconvective vortices. For a very small supercriticality, these
steady vortices behaved sinusoidally along the membrane surface, but with a
small increase in $\Delta V$ above the critical potential drop (about 0.5\% of
$\Delta V_*$), the steady solution lost its sinusoidal profile along the
$x$\nobreakdash-coordinate and acquired a typical spike-like distribution. The
neighboring vortices of opposite circulations carry a significant ion flux (for
the space charge and stream-line distributions see Fig.~\ref{aaa12}(a)). The VC
characteristic dramatically changes for the corresponding $\Delta V$ (see
Fig.~\ref{aaa10}). The basic wave number of these solutions, $k_s$, along with
its overtones, had nonzero Fourier amplitudes; all other Fourier modes which
were not multiples of $k_s$ decayed to zero during the evolution. The value of
the wave number $k_s$ depended upon a particular realization of the
random-number generator \eqref{x1}.

Our next step was to extend the prediction of the weakly nonlinear analysis to
finite supercriticality, namely: a)~to build up a one-parameter family of these
solutions with the parameter~$k_s$; b)~to find the region of stability of that
family.

The problem of the stability of these two-dimensional solutions is much more
difficult than for 1D solutions. If small perturbations are superimposed on the
steady space-periodic solution of this family, these perturbations will be
described by a system with time independent coefficients and will have period
$2\pi/k_s$ in the $x$\nobreakdash-coordinate. According to Floquet's theorem,
the elementary solution of this linear system, bounded at $x=\pm\infty$, has the
form
\begin{equation}\label{L2b}
G(x,y)\cdot\exp{(i\chi x)}\cdot\exp{(\mu t)},
\end{equation}
where $G$ satisfies the corresponding boundary conditions at $y=0$ and $y=1$ and
has the period $2\pi/k_s$ with respect to~$x$, the wave number $\chi$ is a
parameter which is changing within the interval, $0<\chi<k_s/2$, and, hence,
includes side-band and subharmonic bifurcations, and $\mu$ is an eigenvalue of
the problem.

Taking, in the simulations, small enough $k$ (or a large enough domain in~$x$),
we were able to study the problem numerically; $k$ was assumed to be a multiple
of $k_s$, namely, there is an integer number $N$ such that $k\cdot N=k_s$. These
perturbations can be described by the expansion
\begin{equation}\label{GL5}
G = \psi_1(x,y) \cdot e^{\mu_1 t} + \psi_2(x,y) \cdot e^{\mu_2 t} + \psi_3(x,y)
 \cdot e^{\mu_3 t} + \dots,
\end{equation}
where $\psi_k$ are functions with period $2\pi/k$ in~$x$, and $\{\mu_k\}$ are
eigenvalues of the discrete spectrum. Such a stability analysis of
space-periodic solutions for much simpler problems can be found, for example,
in \cite{ChDemKop}. The stability eigenvalues were ordered according to their
real parts,
\[ \Re{(\mu_1)} > \Re{(\mu_2)} > \Re{(\mu_3)} > \dots \]

\begin{figure}[p]
\centering
\includegraphics[angle=-90,width=.9\textwidth]{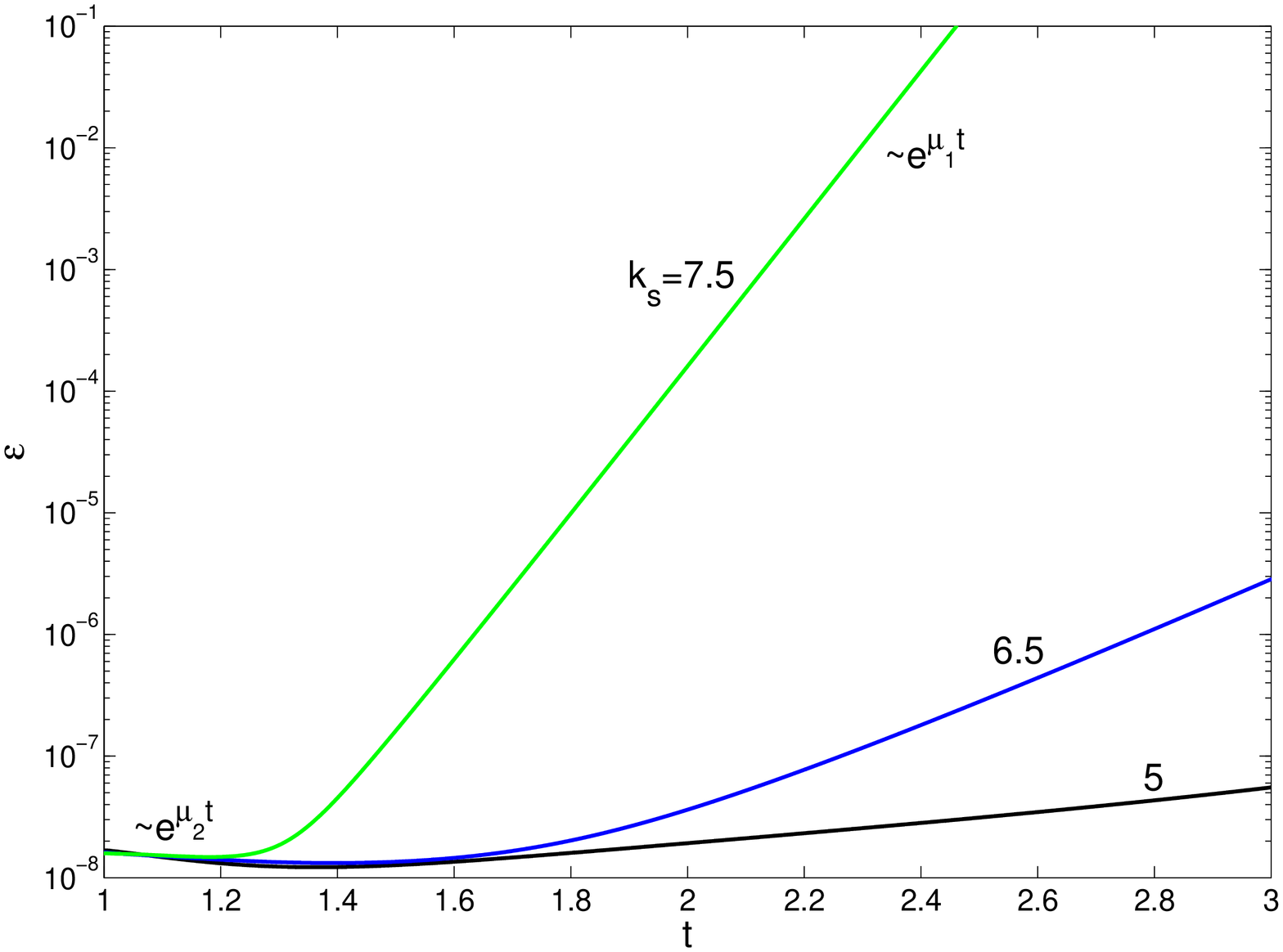}\raisebox{-3.5em}[0em][0em]{\makebox[0em]{\hspace{-3em}(a)}}\\
\includegraphics[angle=-90,width=.9\textwidth]{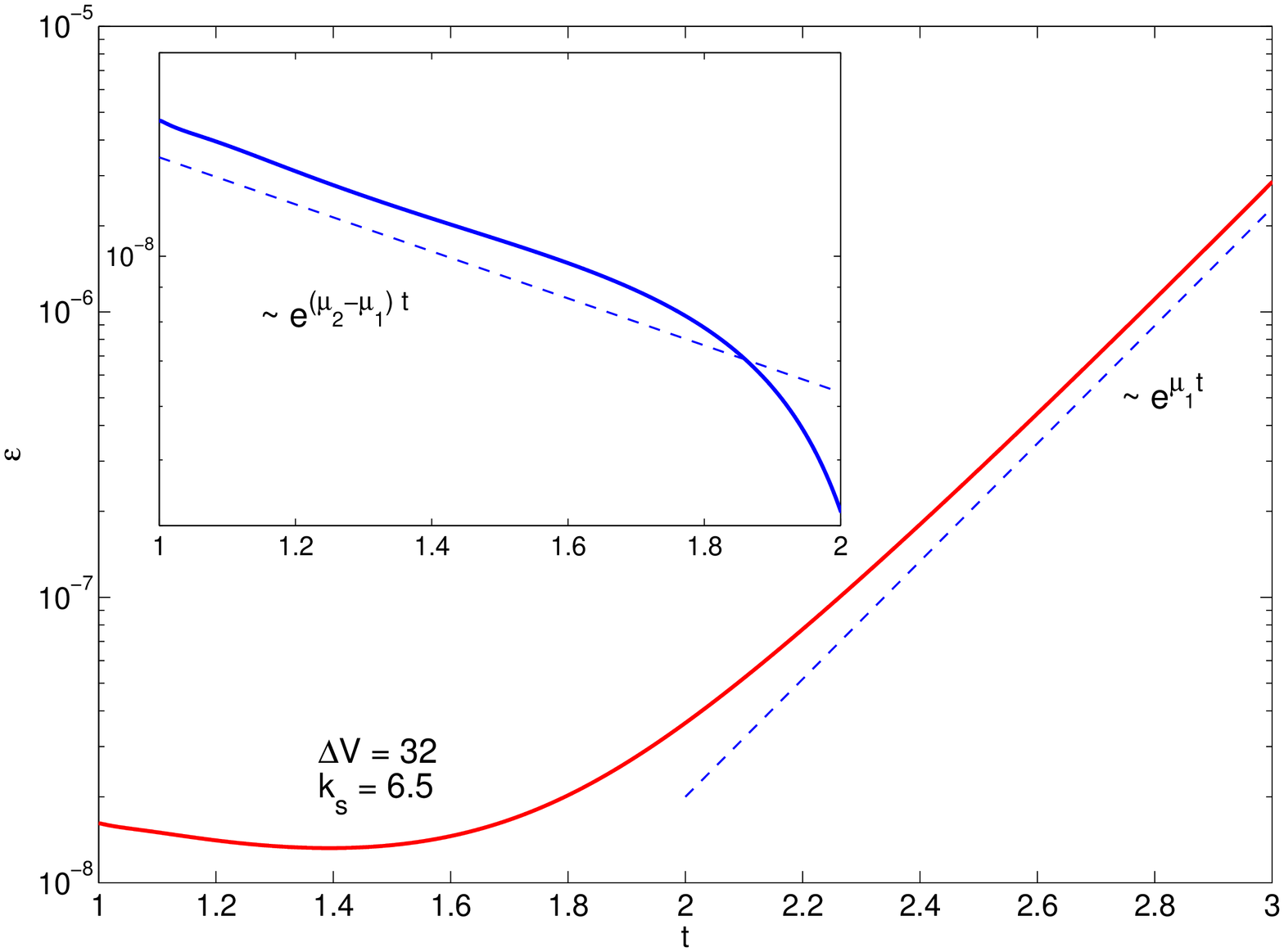}\raisebox{-3.5em}[0em]
[0em]{\makebox[0em]{\hspace{-3em}(b)}}
\caption{(Color online) Amplitude of perturbation vs time, $\varepsilon(t)$,
for different wave numbers $k_s$ (a); scheme for evaluating the first
eigenvalues of stability, $\mu_1$ and $\mu_2$, for $k_s=6.5$ (b);
$\varkappa=0.1$, $\Delta V=34$}
\label{stab}
\end{figure}

\begin{figure}[p]
\centering
\includegraphics[angle=-90,width=.9\textwidth]{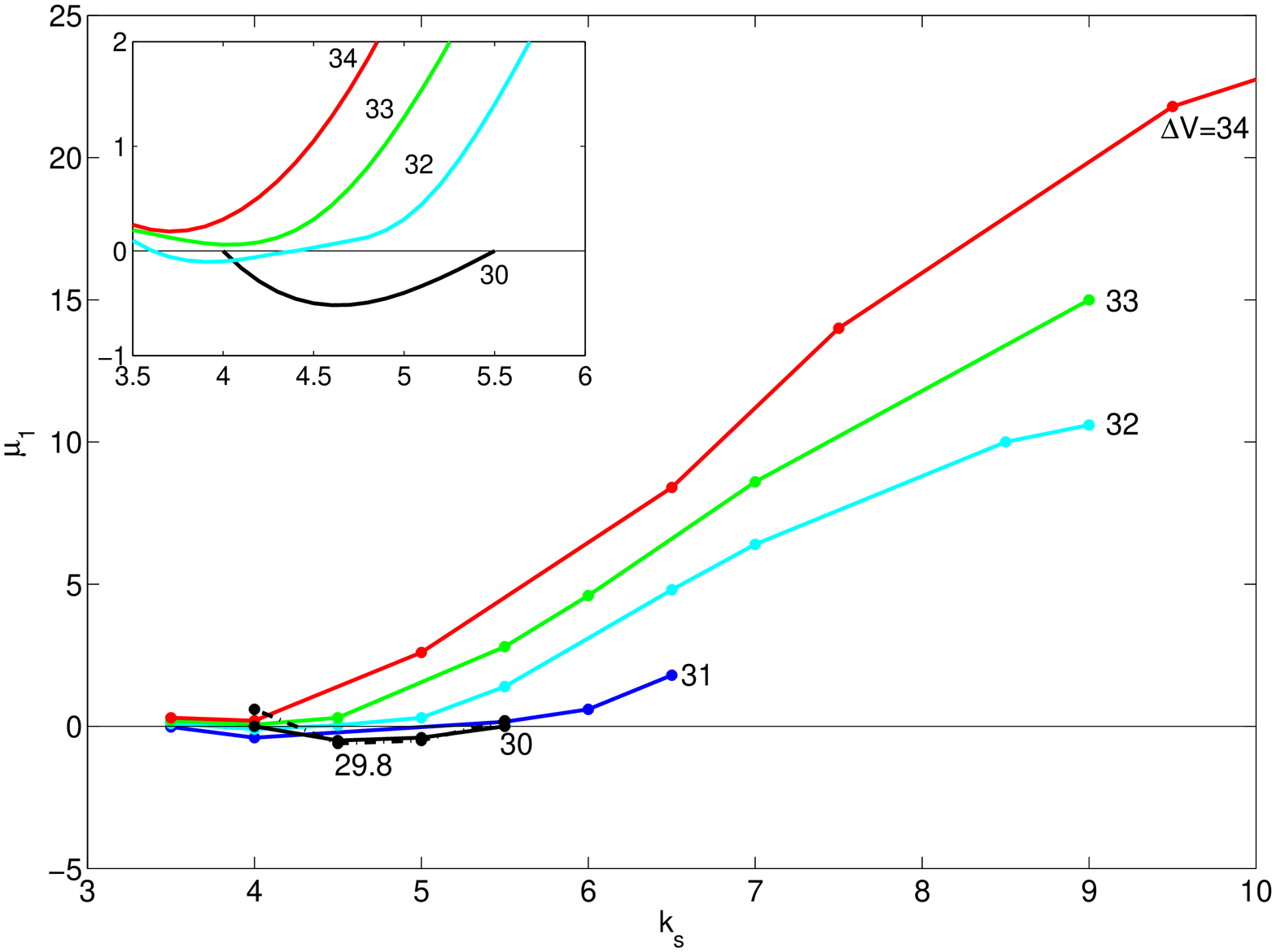}\raisebox{-3.5em}[0em][0em]{\makebox[0em]{\hspace{-3em}(a)}}\\
\includegraphics[angle=-90,width=.9\textwidth]{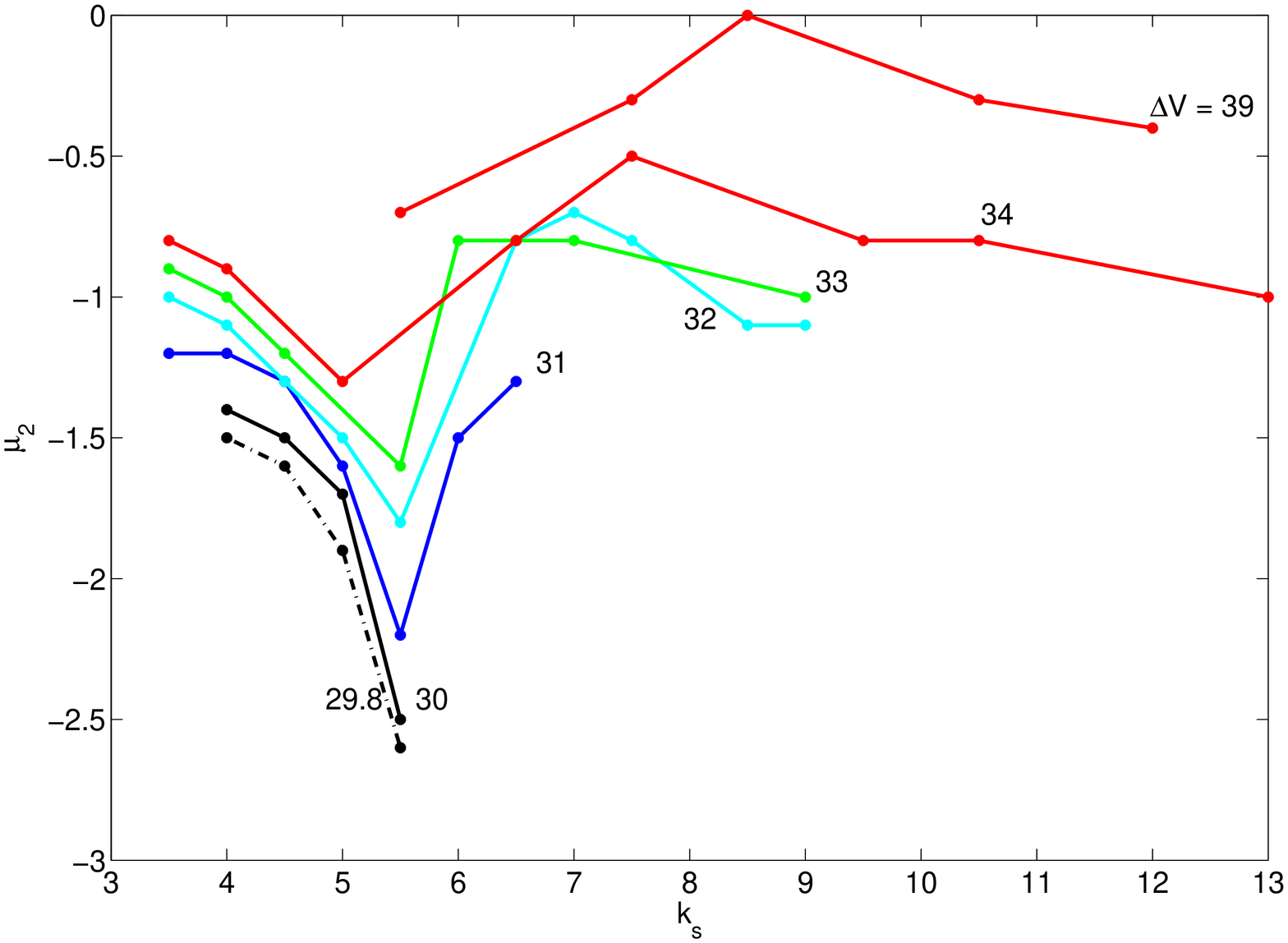}\raisebox{-3.5em}
[0em][0em]{\makebox[0em]{\hspace{-3em}(b)}}
\caption{(Color online) Dependence of the first eigenvalues, $\mu_1$ and
$\mu_2$, on the wave number $k_s$ for different $\Delta V$,
$\varkappa=0.1$}
\label{L1L2}
\end{figure}

The first stability exponents $\mu_1$ and $\mu_2$ were found using the following
algorithm.
\begin{enumerate}\renewcommand{\labelenumi}{\alph{enumi})}
\item The steady 2D solution for a given $k_s$ was calculated by numerically
integrating \eqref{eq1}--\eqref{eqq6} with the initial conditions \eqref{eqq66},
from $t=0$ to some $t=t_0$, until all transients decayed, and we were close
enough to the steady state.
\item The solution found was duplicated $N$ times in the
$x$\nobreakdash-direction, so that its wave number was $k=k_s/N$. The
calculations were restricted by $k=0.5$, which was small enough to find the long
wave instability; the calculations were verified by doubling the
$x$\nobreakdash-domain, $k=0.25$.
\item Small random perturbations $\hat{f}^{\pm}(x,y)$ with $\hat{f}^{\pm} \sim
10^{-8} \div 10^{-6}$ were imposed on $c^+_0$ and $c^-_0$:
\begin{equation}\label{L2c}
t=t_0: \quad c^+=c^+_0 +\hat{f}^+(x,y), \quad c^-=c^-_0+\hat{f}^-(x,y).
\end{equation}
The details of the space distribution of $\hat{f}^{\pm}$ did not affect the
final result.
\item The system \eqref{eq1}--\eqref{eq5} with the initial conditions
\eqref{L2c} was integrated for $t > t_0$ and simultaneously the mean square
amplitude, $\varepsilon(t)$, was computed,
\begin{equation}\label{GL6}
\varepsilon(t)^2 = \frac{k}{2\pi}\int_0^{1}\int_0^{2\pi/k}(c^+-c^+_0)^2
 + (c^--c^-_0)^2 \, dx dy,
\end{equation}
where $c^{\pm}_0$ were kept constant and the $c^{\pm}$ were functions of time.
\end{enumerate}
Substitution of \eqref{GL5} into \eqref{GL6} results in the expansion
\begin{equation}\label{GL7}
\varepsilon(t) = C_1 \cdot e^{\mu_1 t} + C_2 \cdot e^{\mu_2 t} + \dots
 \quad \text{for} \quad t \to \infty,
\end{equation}
where $C_k$ are some constants. The typical behavior of $\varepsilon(t)$ for
different $k_s$ is presented in Fig.~\ref{stab} in semi-logarithmic coordinates,
for large enough times $\log{\varepsilon(t)} \sim \mu_1 t$. The first stability
eigenvalue was obtained as a function of $k_s$; this function, $\mu_1(k_s)$, is
shown in Fig.~\ref{L1L2} for different potential drops $\Delta V$. One can see
that there is a window where the steady periodic solutions are stable. The
boundary of this window is plotted in Fig.~\ref{egg}. For small supercriticality
there is a reasonably good correspondence to the weakly nonlinear analysis. The
stability balloon is finite, its right boundary for $\varkappa=0.1$,
$\Delta V_{**}\approx 32.5$, see Fig.~\ref{egg}. In the region
$\Delta V_{*}<\Delta V<\Delta V_{**}$, the eigenvalue $\mu_1$ is always a real
number.

With the known $\mu_1$ we also managed to calculate the second eigenvalue
$\mu_2$, using the following two-step procedure:
\begin{equation}\label{GL7ab}
\text{(a)} \quad\varepsilon(t)\cdot e^{-\mu_1 t} =C_1+C_2 \cdot
e^{(\mu_2-\mu_1)t}+\dots,
\quad\text{(b)}\quad \varepsilon(t)\cdot e^{-\mu_1 t}
-C_1=C_2 \cdot e^{(\mu_2-\mu_1)t}+\dots
\end{equation}
The procedure for evaluating $\mu_2$ is illustrated in Fig.~\ref{stab}(b),
and the dependence $\mu_2(k_s)$ for different $\Delta V$ is given in
Fig.~\ref{L1L2}. For $\Delta V<39$ all the eigenvalues $\mu_2$ are real and
negative, but for $\Delta V>41$ all $\mu_2$ become positive (not shown in the
figure).

\subsection{Homoclinic contours, their bifurcations, and the birth of chaos}
For $\Delta V > \Delta V_{**}$ all the steady 2D vortices are unstable and the
behavior becomes more complex. In order to clarify the behavior of the system it
was instructive to consider its trajectories in the relevant phase space. The
system has a phase space of high dimension, and it was crucial for its
understanding to choose the proper projection of this space. The electric
current averaged along the membrane, $\langle j\rangle$, and its time
derivative, $d\langle j\rangle/dt$, seemed to be such a representative
low-dimensional projection to characterize the different attractors.


\begin{figure}[p]
\centering
\includegraphics[width=.41\textwidth]{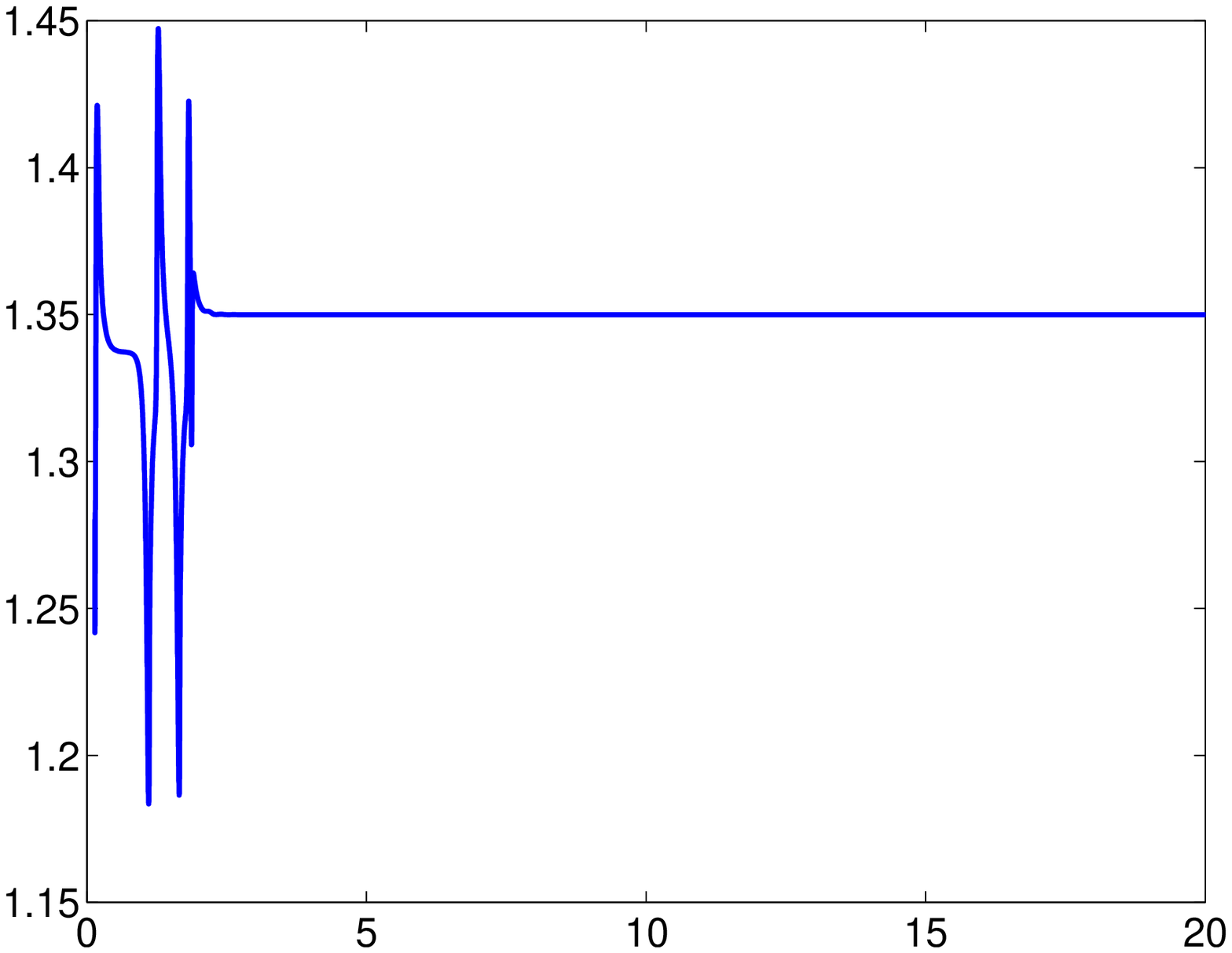}\raisebox{12.5em}[0em][0em]
{\makebox[0em]{\hspace{-38em}a)}}
\includegraphics[width=.41\textwidth]{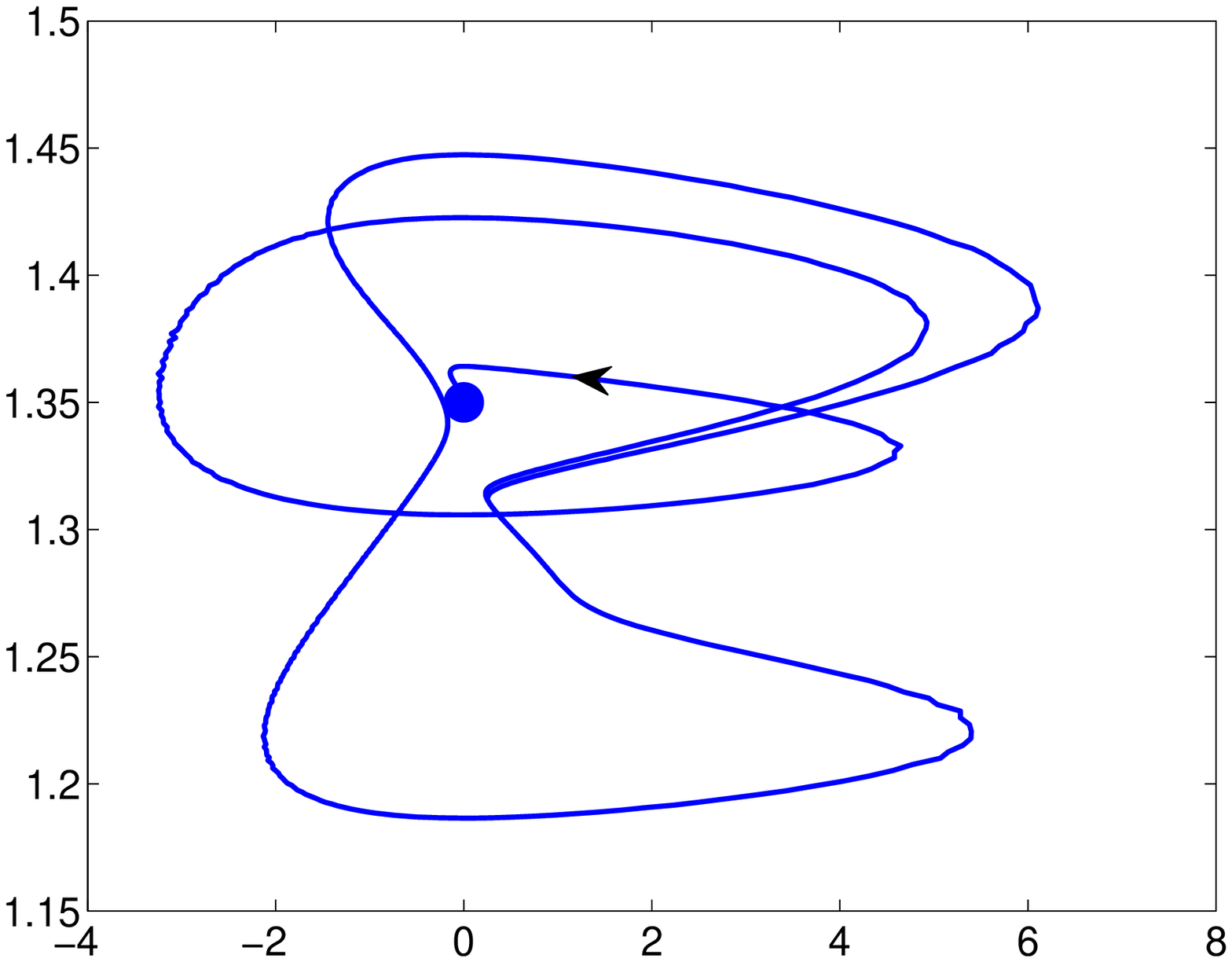}\\
\includegraphics[width=.41\textwidth]{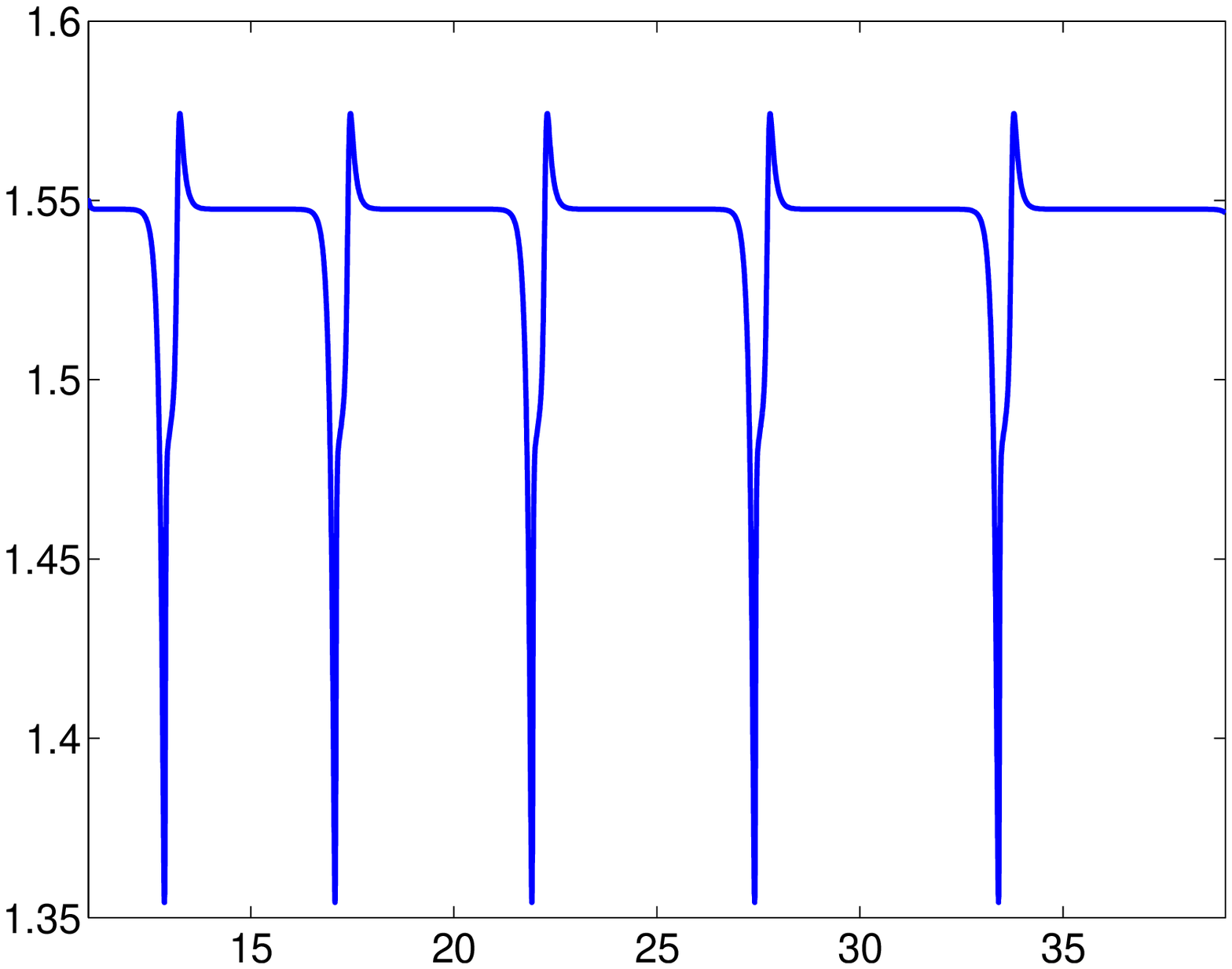}\raisebox{13em}[0em][0em]
{\makebox[0em]{\hspace{-38em}b)}}
\includegraphics[width=.41\textwidth]{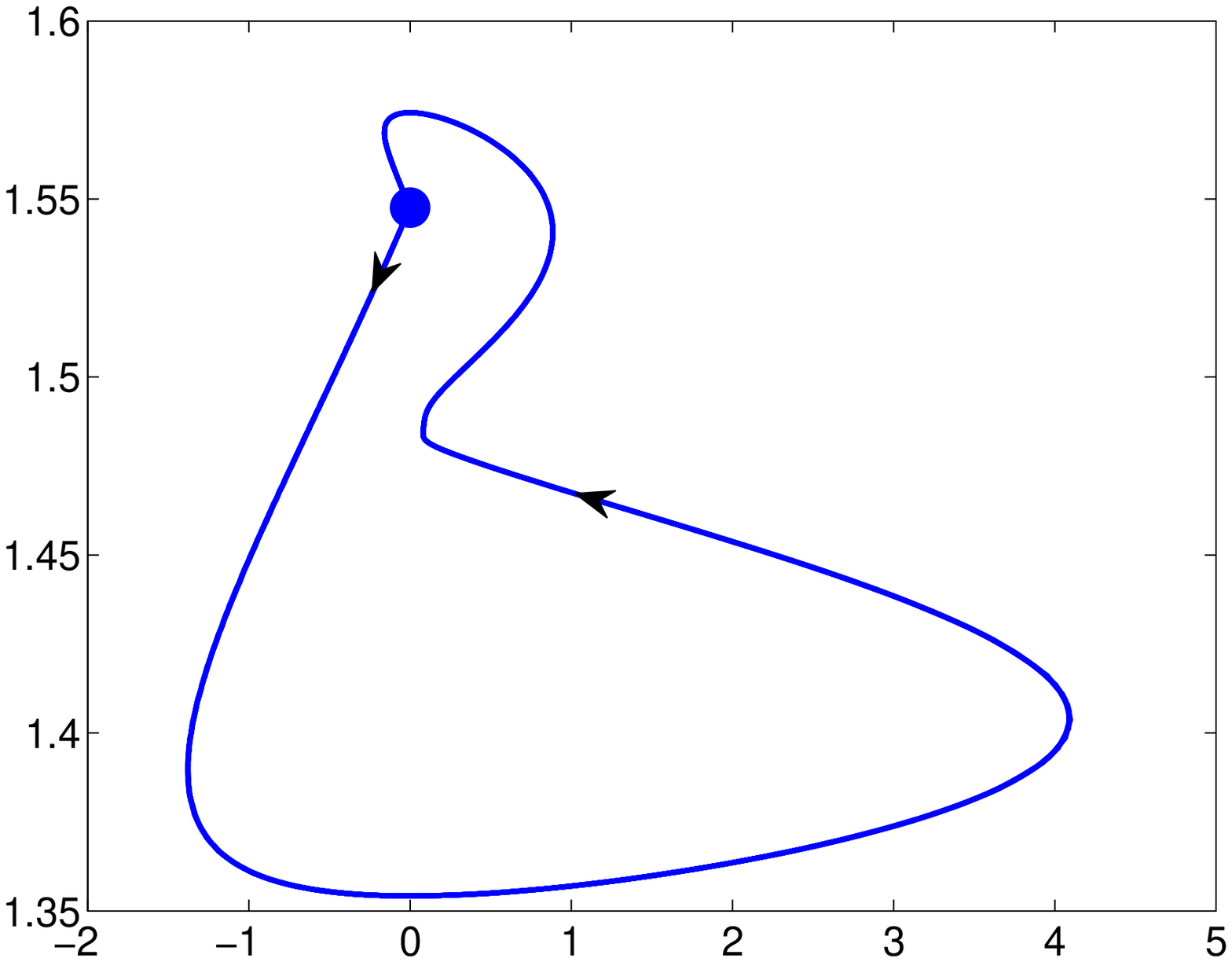}\\
\includegraphics[width=.41\textwidth]{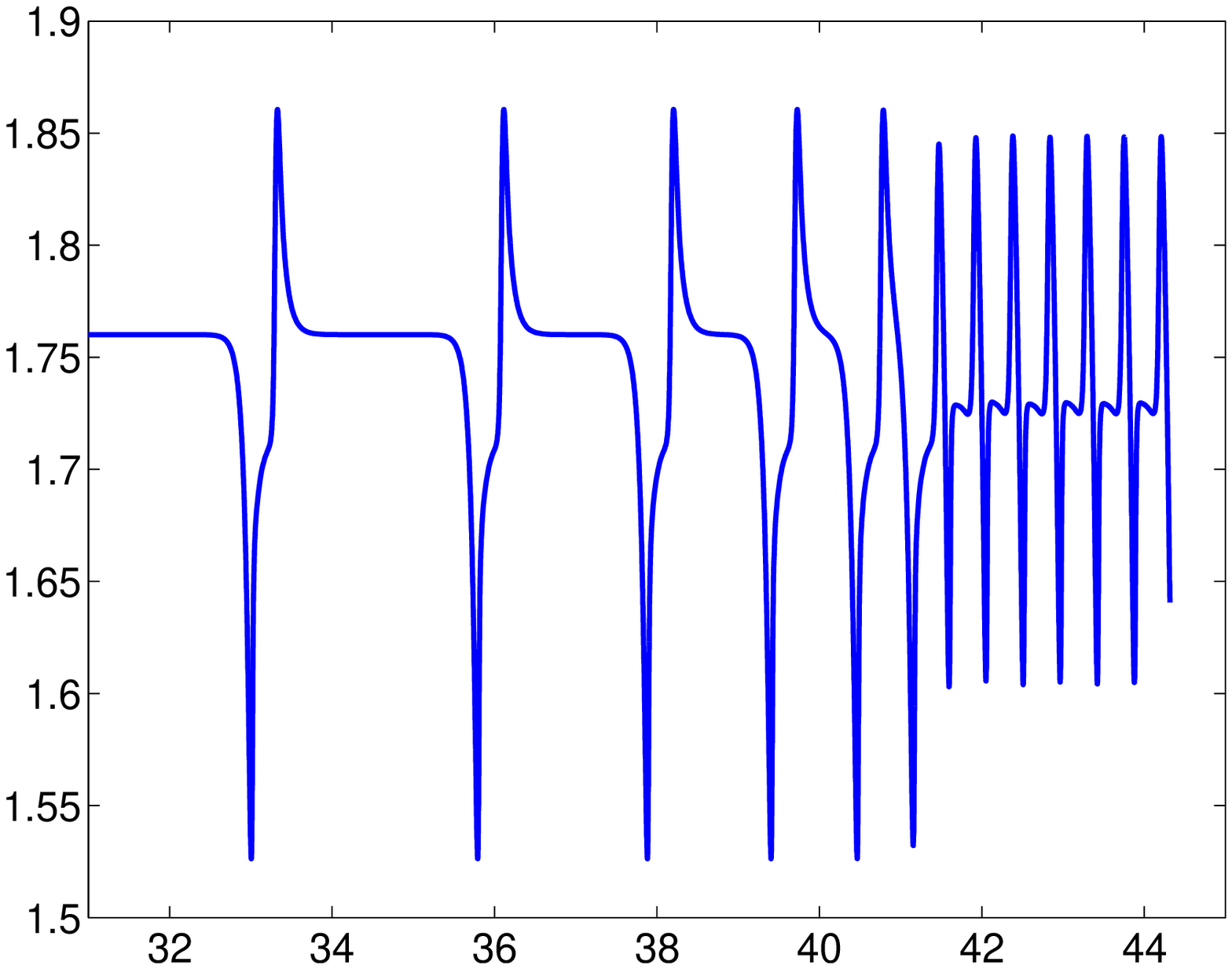}\raisebox{13em}[0em][0em]
{\makebox[0em]{\hspace{-38em}c)}}
\includegraphics[width=.41\textwidth]{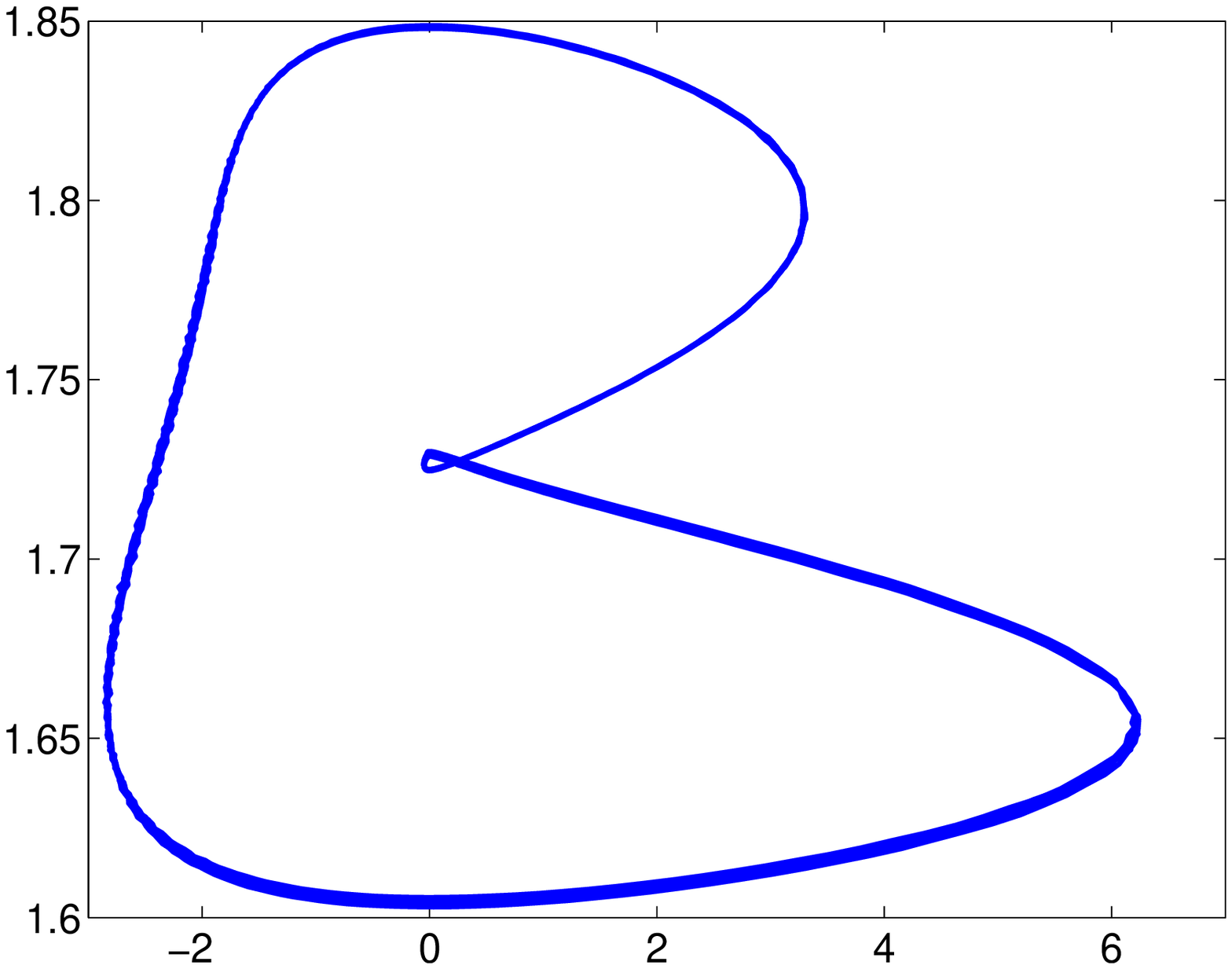}\\
\includegraphics[width=.41\textwidth]{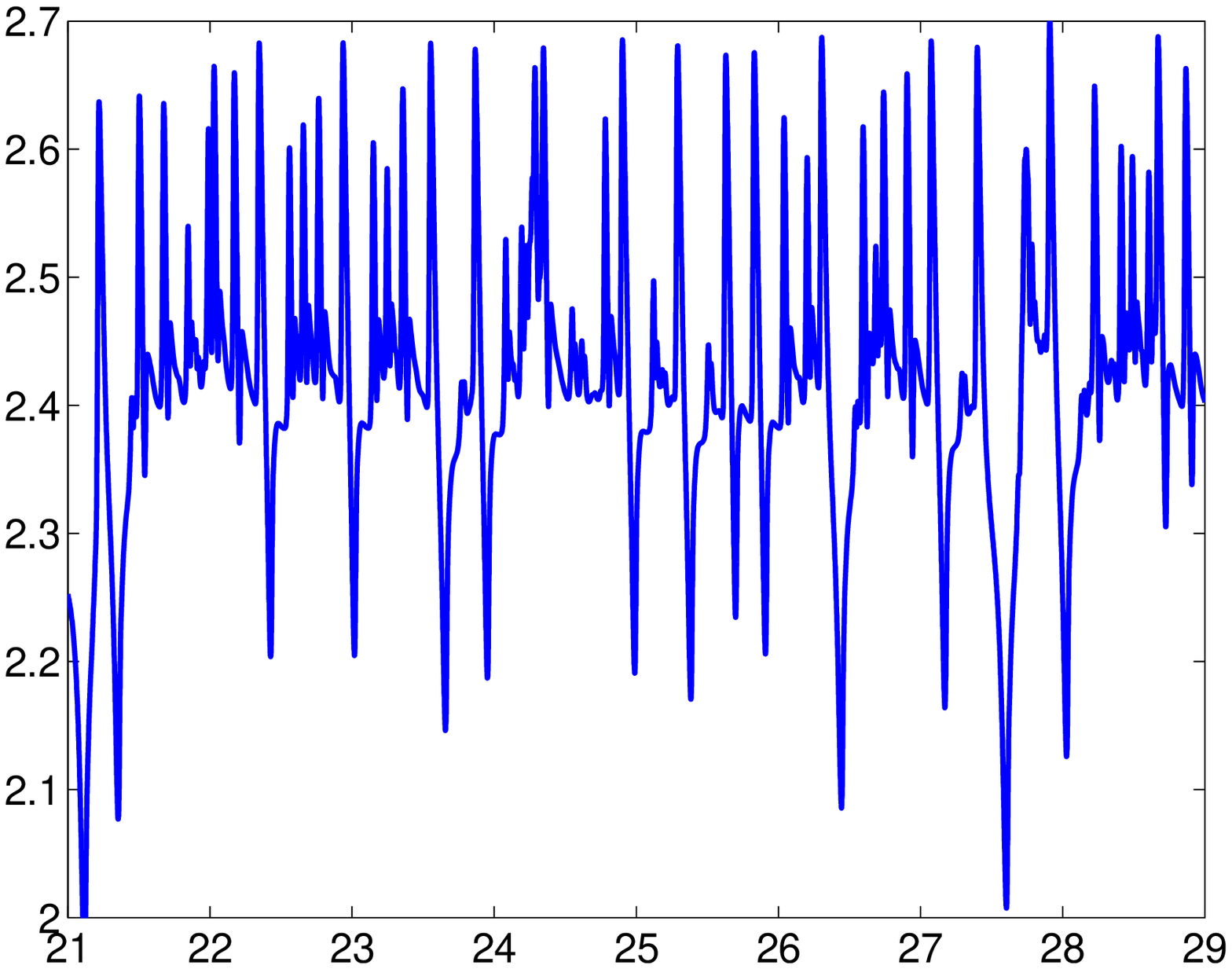}\raisebox{13em}[0em][0em]
{\makebox[0em]{\hspace{-38em}d)}}
\includegraphics[width=.41\textwidth]{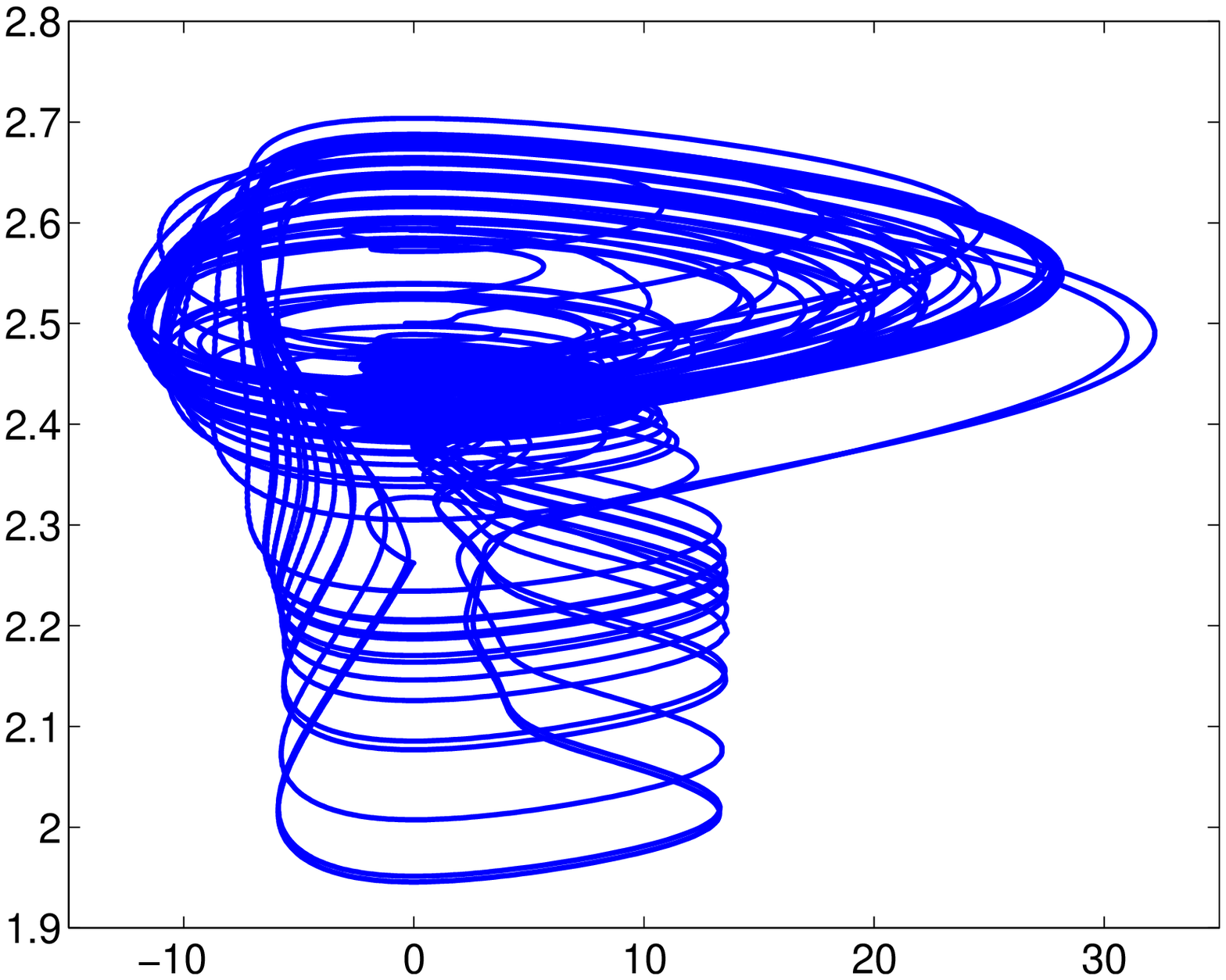}
\caption{(Color online) Time evolution and phase diagram for the average
electric current: a) $\Delta V=31$; b) $\Delta V=33$; c) $\Delta V=35$, d)
$\Delta V=40$. For all the calculations $\varkappa=0.1$}
\label{aaa13}
\end{figure}

For the regime of steady 2D vortices, $\Delta V_* <\Delta V < \Delta V_{**}$,
the phase trajectory at $t \to \infty$ is attracted to some fixed point,
$\langle j\rangle=\const$ and $d\langle j\rangle/dt=0$, with wave number $k_s$
from the stability interval, see Fig.~\ref{aaa13}(a). With increasing
$\Delta V$, this stability interval narrows, and at $\Delta V = \Delta V_{**}$
shrinks to one point, for $\varkappa=0.1$ and $\nu=10^{-3}$, $k_s\approx 4$.

For $\Delta V > \Delta V_{**}$ the stability interval vanishes. All the
stationary points become saddle-node points, making a one-dimensional unstable
manifold which corresponds to the real eigenvalue $\mu_1>0$. After a long
transition period there was established the behavior: (a)~the phase trajectory
was rapidly attracted to some saddle point along its stable manifold; (b)~the
solution spent a significant time near this stationary state; (c)~the solution
was rapidly expelled from the vicinity of the saddle point along its
one-dimensional unstable manifold (d)~with a subsequent return to the stationary
saddle point, etc. The attracting--repelling scenario ran over many times.
Return to the steady state was aperiodic; and with each new cycle, the time
spent in the steady state increased. Apparently, the attractor of the system was
a homoclinic contour, consisting of the saddle fixed point and trajectories
doubly asymptotic to this point as $t \to \pm \infty$ (see, for example,
\cite{GH}). The typical projections are shown in Fig.~\ref{aaa13}(b).

In the physical plane the behavior can be described as follows: the solution for
a long time looked like a steady 2D electroconvective vortex; after this time
the neighboring spikes of the space charge either merged quickly to create
larger ones or disintegrated and broke up into smaller ones with a final return
to the quasi-steady structure, etc. (see Fig.~\ref{aaa12}(b)). The wave number
of this quasi-steady structure was $k_s \approx 2 \div 4$.

The behavior for simpler dynamical systems with a saddle-node point is well
known, see \cite{GH}. The dynamics in the phase space of such a system is
dominated by the saddle point, because the phase trajectory spends most of its
time near this point. $S= \mu_1 +\mu_2$ ($\mu_1>0$ and $\mu_2 <0$) is called the
saddle value. If $S$ is negative, this means that the saddle point ``attracts
stronger'' than it ``repels''. Then the attractor of the system is a homoclinic
contour which consists of the saddle point and the loop which is the
intersection of the stable and unstable manifolds of the saddle point. If $S$ is
positive, this contour becomes unstable and what results from bifurcation from
the homoclinic orbit limit cycle inherits stability.

\begin{figure}[hbtp]
\centering
\includegraphics[width=\textwidth]{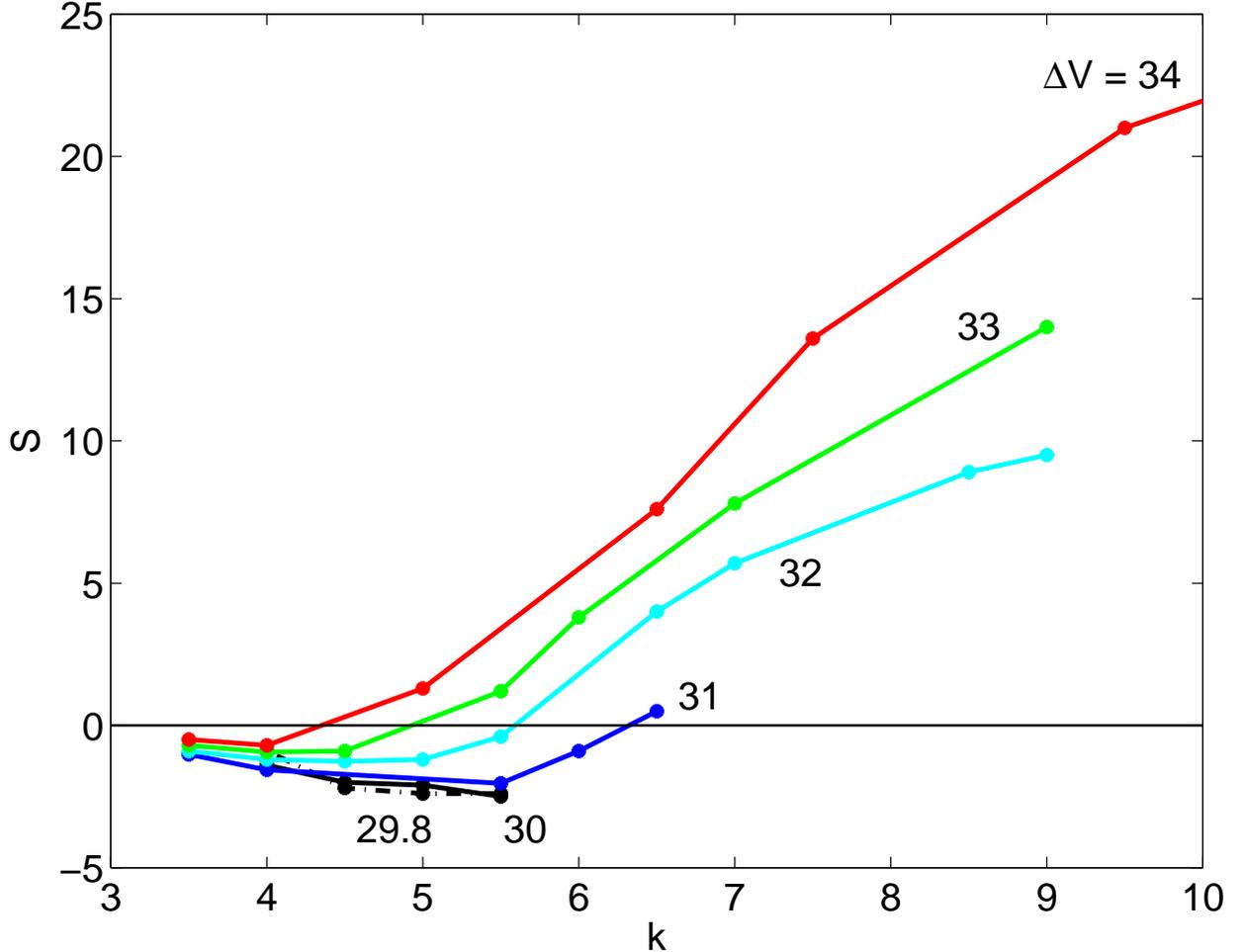}
\caption{(Color online) Saddle value $S$ as a function of $k_s$ for different
potential drops, $\varkappa=0.1$}
\label{s}
\end{figure}

Our problem is more complex, but its behavior seems to preserve the main
features of the finite dynamical system with a homoclinic contour. We calculated
the saddle value $S$ for $\Delta V > \Delta V_{**}$ and presented it in
Fig.~\ref{s} as a function of the wave number $k_s$ for different potential
drops. The boundary with $S=0$ is denoted in Fig.~\ref{egg} by the line~3. The
figure shows that for $\Delta V < \Delta V_{***}\approx 37$ there are saddle
points with negative~$S$, a fact which we interpret as the existence of stable
homoclinic contours.
With increasing $\Delta V$, $\Delta V>\Delta V_{***}$, $S$ became positive and
the contour became unstable. Numerical simulations of \eqref{eq1}--\eqref{eq5}
with the white noise initial conditions \eqref{eqq67} confirmed the birth of a
stable cycle. This is illustrated in Fig.~\ref{aaa13}(c) where it is shown that
the limiting solution acquired periodic oscillations in time. With a further
very small increase of $\Delta V$, by about 2\% of $\Delta V_*$, the limit cycle
lost its stability and the oscillations became aperiodic and chaotic, see
Fig.~\ref{aaa13}(d). In this research we did not investigate the details of the
transition from periodic to chaotic motion.

With larger $\Delta V$, the oscillations looked much more chaotic, for the
behavior in the phase space see Fig.~\ref{aaa13}(d). The spikes of the space
charge not only coalesced aperiodically but their coalescence became very
disordered and provided high speed outflow jets upwards (see
Fig.~\ref{aaa12}(c)).

\subsection{Spatial spectra}
\begin{figure}[hbtp]
\centering\begin{tabular}{@{}ll@{}}
\qquad (a) & \qquad (b)\\
\includegraphics[width=.49\textwidth]{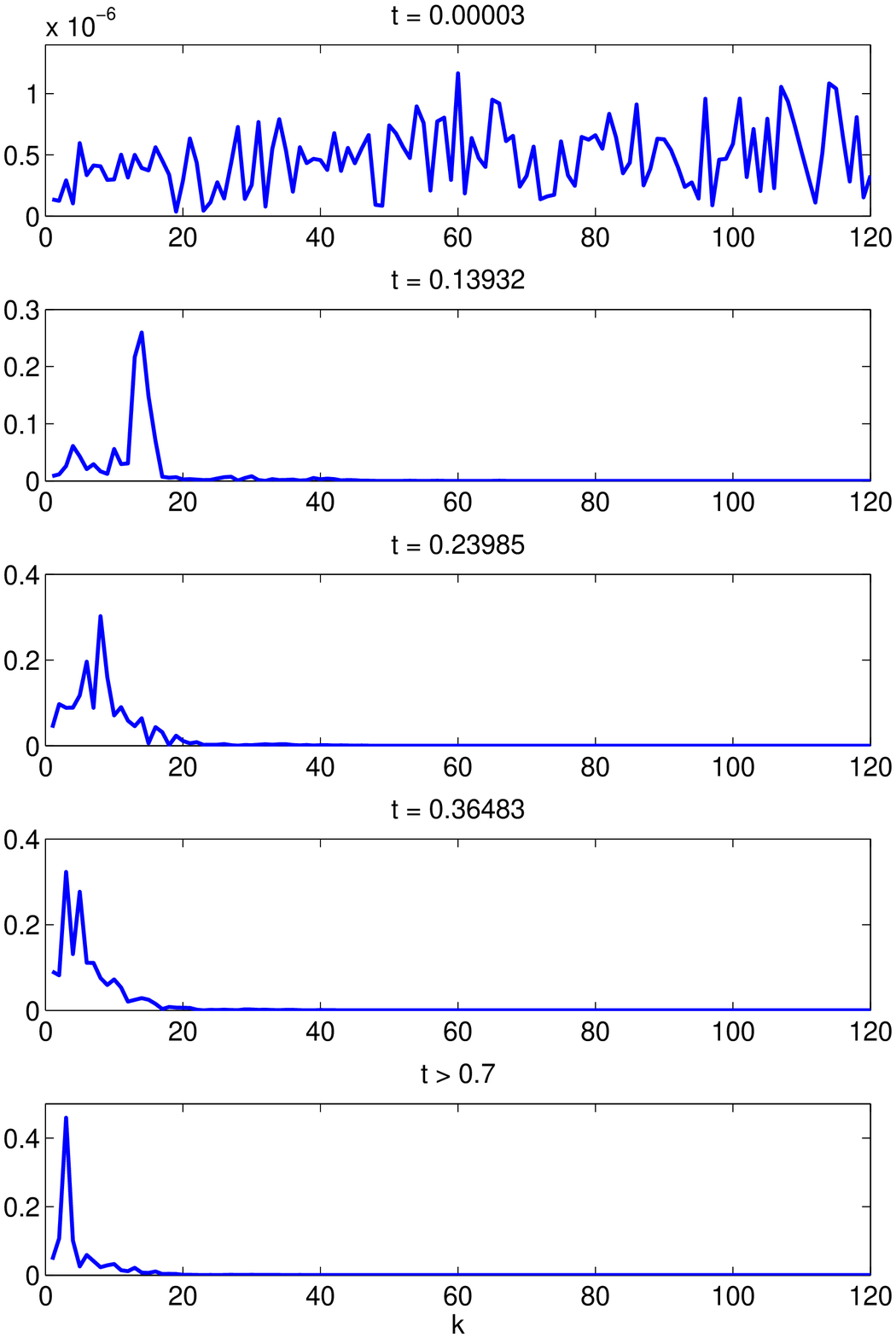}
 & \includegraphics[width=.49\textwidth]{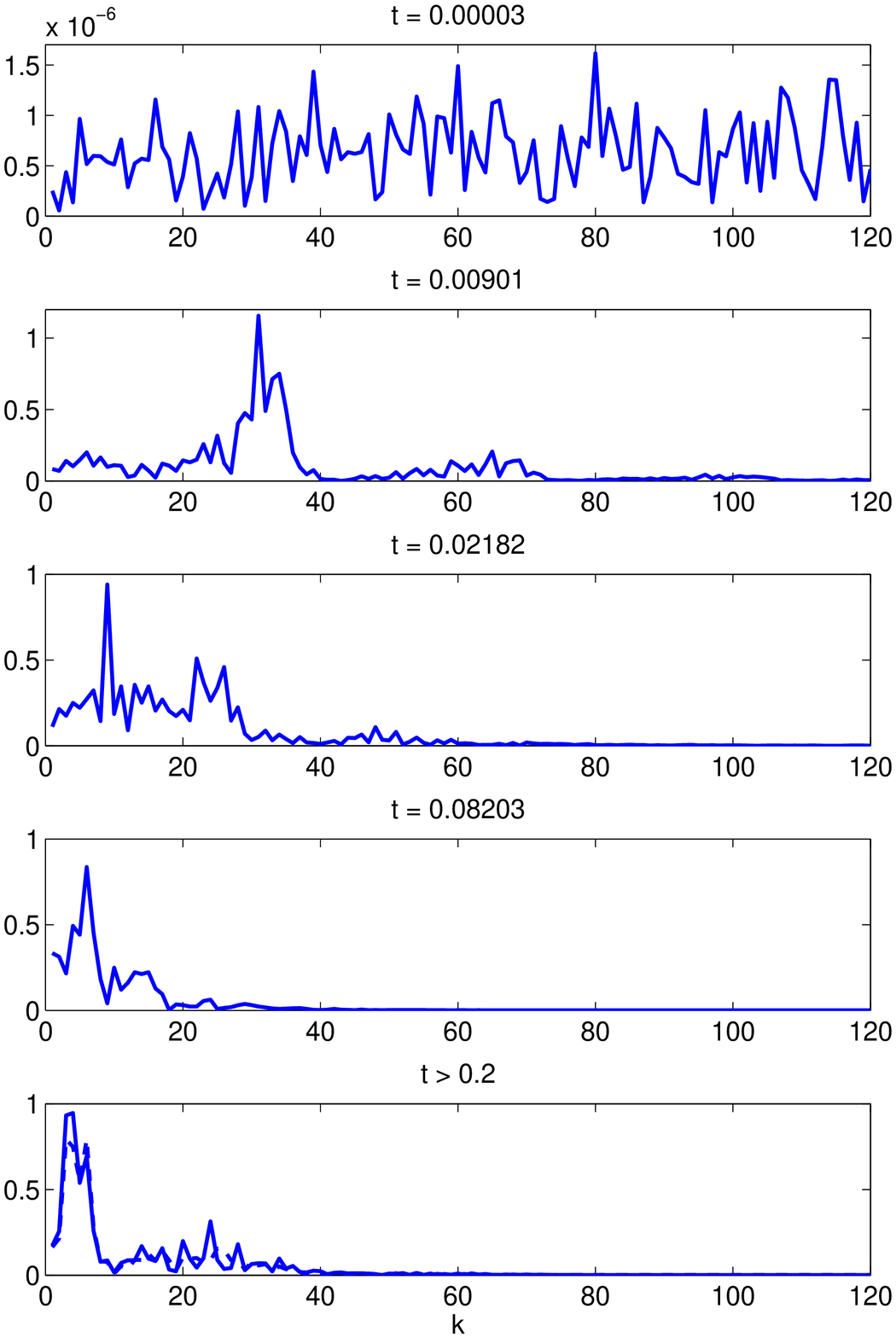}
\end{tabular}
\caption{(Color online) Evolution of the spatial Fourier transform $F(k)$,
$\varkappa=0.1$ (a) $\Delta V=30$ and (b) $\Delta V=50$}
\label{aaa21}
\end{figure}

A~spatial Fourier transform of the electric current $j(x)$,
\begin{equation}\label{q1}
F(k) = \left|\int_0^{\infty} j e^{-i k x} dx \right|,
\end{equation}
gives the characteristic size and characteristic wave number of the vortex at
any instant of time, $t>0$. In Fig.~\ref{aaa21}(a), the evolution of $F(k)$
within the interval $\Delta V_* <\Delta V < \Delta V_{**}$ is shown. The
small-amplitude initial white noise ($t=3\times 10^{-5}$) after filtering by the
linear instability evolves into a narrow primary band with a characteristic wave
number $k \approx 13$ ($t=1.4\times 10^{-1}$. Beyond this time, however,
nonlinearity corrupts this spectrum and wave coalescence makes the
characteristic wave length longer, with eventual wave number $k_D \approx 5$.

In Fig.~\ref{aaa21}(b), the typical spectrum evolution for the chaotic regime
is presented. After the linear filtering, at $t=9 \times 10^{-3}$, overtone and
subharmonic bands appear. At the final stage of evolution, the perturbations
again acquire coherency, now due to nonlinearity, the characteristic wave number
$k_D \approx 5$ but it is not steady and it oscillates in time within a narrow
frequency window (dashed line) and its Fourier distribution has a pronounced
high-frequency band which is also oscillating.
For all our calculations, the typical characteristic wave number of the
attractor is within the interval $k_D = 3 \div 5$. It is close to the critical
wave number of the instability threshold $k_*$, see Table~\ref{T3}. The typical
length of the attractor pattern is a multiple of from 1.25 to 2 of the distance
between the membranes,~$\tilde{L}$.

\begin{table}
\centering
\caption{Dominant wave number for $\nu=10^{-3}$ and (a)~$\varkappa=0.2$;
(b)~$\varkappa=0.1$; (c)~$\varkappa=0.05$}\label{T4}
\begin{tabular}{|c|c|}
\hline
\multicolumn{2}{|c|}{(a)}\\
\hline
$\Delta V$ & $k_D$ \\
\hline
28& 5.0\\
\hline
30& 4.0\\
\hline
32& 3.0\\
\hline
35&3.0\\
\hline
36&3.0\\
\hline
37&4.0\\
\hline
38&5.0\\
\hline
39&5.0\\
\hline
40&4.0\\
\hline
50&4.0\\
\hline
\end{tabular}
\qquad
\begin{tabular}{|c|c|}
\hline
\multicolumn{2}{|c|}{(b)}\\
\hline
$\Delta V$ & $k_D$ \\
\hline
32& 5.0\\
\hline
33& 4.0\\
\hline
33.5& 4.0\\
\hline
34&4.0\\
\hline
34.7&3.0\\
\hline
34.8&3.0\\
\hline
35&3.0\\
\hline
37&3.0\\
\hline
40&3.0\\
\hline
50&4.0\\
\hline
\end{tabular}
\qquad
\begin{tabular}{|c|c|}
\hline
\multicolumn{2}{|c|}{(c)}\\
\hline
$\Delta V$ & $k_D$ \\
\hline
40& 4.0\\
\hline
42& 4.0\\
\hline
44& 3.0\\
\hline
46&3.0\\
\hline
48&2.0\\
\hline
50&2.0\\
\hline
52&2.0\\
\hline
54&4.0\\
\hline
56&4.0\\
\hline
60&4.0\\
\hline
\end{tabular}
\end{table}

The simulations were done for the doubled interval, $k=0.5$; the spectrum did
not change. The behavior for different $\varkappa$ is qualitatively identical
but with increasing $\varkappa$ the solution becomes more chaotic. The dominant
wave numbers for $\nu=10^{-3}$ and different $\varkappa$ are presented in
Table~\ref{T4}.

\subsection{Comparison with experiment}
Electrokinetic instability is a new type of electro-hydrodynamic instability for
which experimental evidence was presented only recently
(\cite{Belv,RubRub1,YCh,KmHn}). Direct quantitative comparison with experimental
data is complicated by many reasons, maybe the main reason is that it is
difficult to take into account all the experimental factors involved. Still we
will try to evaluate the experimental and theoretical data.

Only the paper \cite{Belv} contains time series of variables
(chronopotentiometric curves) and, in qualitative agreement with our
simulations, experimental time series do not show regular periodic oscillations,
sinusoidal or close to sinusoidal: there is no evidence of Hopf bifurcation in
experiments.

\begin{figure}[hbtp]
\centering
 \includegraphics[height=.26\textheight]{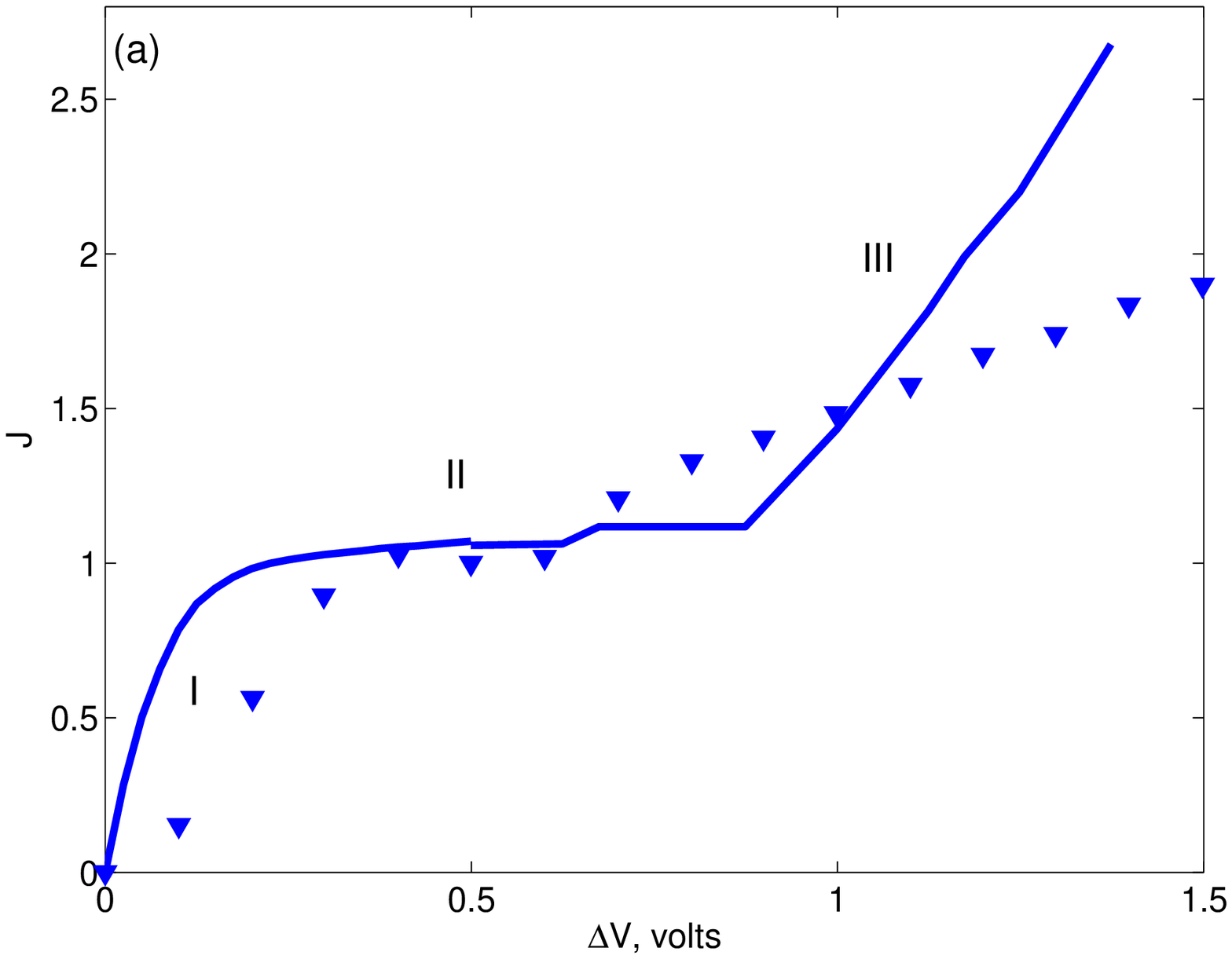}
 \includegraphics[height=.26\textheight]{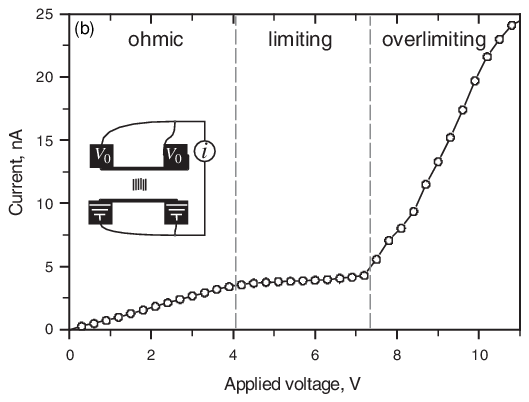}
\caption{(Color online) (a)~VC characteristics for $\kappa=0.05$ and
$\nu=10^{-3}$; I, II and III stand for underlimiting, limiting and overlimiting
currents, respectively; the triangles corresponds to experiments \cite{RubRub1}
and the solid line corresponds to our simulations. (b)~Experimental data from
\cite{KmHn} clearly shows the regions of underlimiting, limiting and
overlimiting currents}
\label{aaa00}
\end{figure}

From the Table~\ref{T4}, taking into account $\varkappa=0.05$ and
$\tilde{L}=0.5$~mm, the dimensional dominant wave number changes within the
window $\tilde{k}_D=4 \div 8 \text{ mm}^{-1}$. This is in qualitative agreement
with $\tilde{k}_D=2 \text{ mm}^{-1}$ in \cite{RubRub1}. The experimental
threshold of instability according to \cite{RubRub1}
$\Delta \tilde{V}_* = 0.5$~V, and the theoretical one is (see Table~\ref{T3})
$\Delta \tilde{V}_* = 0.95$~V.

The experimental \cite{RubRub1} and calculated VC characteristics are presented
in Fig.~\ref{aaa00}(a); again they show a qualitative agreement between the
theoretical and experimental data. The experimental VC behavior obtained in
\cite{KmHn} qualitatively fits the theoretical one in Fig.~\ref{aaa10}
(relatively long region of the limiting currents, rather steep VC-dependence for
the overlimiting region in comparison with the ohmic region). Unfortunately,
many important experimental parameters which are necessary for the comparison
are not presented in \cite{KmHn}.

The final resum\'e is: there is a qualitative agreement between the theoretical
and experimental data. For better quantitative comparison, additional factors
must be taken into account.

\section{Conclusions}

The results of direct numerical simulation of electrokinetic instability,
described by the system of the \NPPS\ equations, were presented. Two kinds of
initial conditions were applied: (a)~white noise initial conditions were taken
to mimic ``room disturbances'' and the subsequent natural evolution of the
solution; (b)~an artificial monochromatic ion distribution with a fixed wave
number was used to create steady coherent structures.

The results were discussed and understood from the viewpoint of bifurcation
theory and the theory of hydrodynamic instability. The following sequence of
attractors and bifurcations was observed as the drop of potential between the
selective surfaces increased: stationary point --- homoclinic contour --- limit
cycle --- chaotic motion; this sequence does not include Hopf bifurcation.

The threshold of electroconvective movement, $\Delta V = \Delta V_{*}$, and the
origin of electroconvective vortices were found by the linear spectral stability
theory which results were confirmed by the numerical simulation of the {\NPPS}
system. For a small supercriticality the electroconvective vortices and their
stability were described by a weakly nonlinear theory. Direct numerical
integration of the entire system extended the results of the weakly nonlinear
analysis to finite supercriticality. In particular, a ``balloon'' of stability
of steady vortices was found: $\Delta V_{*}< \Delta V < \Delta V_{**}$. For a
large enough drop of potential, $\Delta V > \Delta V_{**}$, all the steady
vortices become unstable via real eigenvalues. The new attractor is
characterized by aperiodic oscillations when the solution for a long time looks
like a steady 2D electroconvective vortex, but eventually the neighboring spikes
of the space charge either rapidly coalesce or disintegrate to return to the
electoconvective quasi-steady structure. Such a scenario repeats itself many
times, forming in the proper phase space a so called stable homoclinic contour.

With further increase of $\Delta V > \Delta V_{***}$, the homoclinic contour
loses its stability and the motion first becomes periodic and then, with a very
small further increase of $\Delta V$, chaotic. Numerical resolution of the thin
concentration polarization layer showed spike-like charge profiles along the
membrane surface. For a large enough potential drop, the coalescence of the
spikes of the space charge became very disordered and provided high-speed
outflow jets and the additional ion flux was dominated by these outflow jets.

The numerical investigation confirms the absence of regular, close to
sinusoidal, oscillations for overlimiting regimes. There is a qualitative
agreement between the experimental and theoretical thresholds of instability,
dominant wave numbers of the observed coherent structures, and voltage--current
characteristics.

\begin{acknowledgments}
This work was supported, in part, by the Russian Foundation for Basic
Research (Projects No. 11-08-00480-a, No. 11-01-00088-a and No.
11-01-96505-r\_yug\_ts).
\end{acknowledgments}


\renewcommand{\refname}{References}

\end{document}